\documentclass[aps, twocolumn, reprint, pra,
superscriptaddress,
footinbib]{revtex4-2}

\usepackage{layout}

\usepackage{amsmath,amssymb,bm}
\usepackage{graphicx}
\usepackage{latexsym}
\usepackage[normalem]{ulem} 
\usepackage[usenames,dvipsnames]{color}
\usepackage{hyperref}
\usepackage{natbib}
\usepackage{xcolor}
\usepackage{physics}
\usepackage{stackengine}

\definecolor{grey}{RGB}{165,165,165}
\definecolor{b}{HTML}{1f77b4}
\definecolor{g}{HTML}{2ca02c}
\definecolor{o}{HTML}{ff7f0e}
\definecolor{r}{HTML}{d62728}

\hypersetup{
	colorlinks,
	linkcolor={blue},
	citecolor={blue},
	urlcolor={blue}
}

\usepackage[colorinlistoftodos]{todonotes}

\usepackage{framed}

\usepackage[utf8]{inputenc} 

\newcommand{\ind}[1]{_\text{#1}}

\begin{document}

\title{Single-particle localization in a two-dimensional Rydberg spin system}

\date{\today}

\author{Jan Philipp Klinger}
\affiliation{Kirchhoff-Institut f\"{u}r Physik, Universit\"{a}t Heidelberg, Im Neuenheimer Feld 227, 69120 Heidelberg, Germany}
\author{Martin G\"{a}rttner}
\affiliation{Physikalisches Institut, Universit\"at Heidelberg, Im Neuenheimer Feld 226, 69120 Heidelberg, Germany}
\affiliation{Kirchhoff-Institut f\"{u}r Physik, Universit\"{a}t Heidelberg, Im Neuenheimer Feld 227, 69120 Heidelberg, Germany}
\affiliation{Institut f\"ur Theoretische Physik, Ruprecht-Karls-Universit\"at Heidelberg, Philosophenweg 16, 69120 Heidelberg, Germany}

\begin{abstract}
We study excitation transport in a two-dimensional system of randomly assembled spins with power-law hopping in two dimensions. This model can be realized in cold atom quantum simulators with Rydberg atoms. In these experiments, due to the Rydberg blockade effect, the degree of disorder in the system is effectively tunable by varying the spin density. We study dynamics and eigenstate properties of the model as a function of disorder strength and system size and discuss potential limitations for experiments. At strong disorder we predominantly observe  localized eigenstates with power-law tails. In this regime the spectral and eigenstate properties can be understood in a perturbative picture of states localized on small clusters of spins. As the disorder strength is weakened eigenstates become increasingly delocalized and a set of seemingly multifractal states appears in the low-energy tail of the spectrum. A detailed study of the system-size scaling of the eigenstate properties indicates that in the infinite-size limit all states eventually become localized. We discuss the feasibility of observing localization effects experimentally in the spatial spreading of an initially localized excitation and identify limited system sizes and finite decoherence rates as major challenges. Our study paves the way towards an experimental observation of localization effects in Rydberg spin systems with tunable disorder.
\end{abstract}

\maketitle  

\section{Introduction}
\label{sec:intro}

Anderson localization, the absence of transport of a particle through a disordered medium, is due to interference between different paths that a particle can take, and thus occurs in isolated quantum systems \cite{Anderson1958, Kramer1993}. It has been studied theoretically and observed experimentally in one
\cite{Billy2008, Roati2008} and three \cite{Chabe2008} spatial dimensions using cold atoms and recently also in two-dimensional systems \cite{Manai2015, White2019}, which present a challenge due to the notoriously large localization length. In the original tight-binding Anderson model the particle can hop to its neighboring sites in a lattice geometry and each lattice site has a random potential energy. On the contrary, the case of purely off-diagonal disorder, i.e., random hopping strength and no disorder potential, in combination with power-law interactions is less well studied theoretically \cite{Levitov1990, Lee1981}, but is also of great interest as it is relevant for transport processes in biological systems such as light harvesting complexes \cite{Engel2007, Anderson1998, Walschaers2016}. Moreover, many quantum simulation platforms naturally feature power-law interactions, which combined with randomness in the particle positions, allow the realization of random hopping models. Examples for such experimental implementations include magnetic atoms \cite{dePaz2013, Baier2016}, polar molecules \cite{Yan2013}, trapped ions \cite{Richerme2014, Jurcevic2014, Britton2012}, nitrogen vacancy centers in diamonds \cite{Waldherr2014}, nuclear spins in solid-state systems \cite{Alvarez2015}, atoms trapped in a photonic crystal waveguide \cite{Hung2016}, and Rydberg atoms \cite{Saffman2010}, the last being the main targeted system in this work.

Here we study a power-law Euclidean model with purely off-diagonal disorder in two dimensions (2D), i.e., a particle hopping between randomly placed sites with a hopping strength $V_{ij}\propto r_{ij}^{-a}$, where $r_{ij}$ is the interparticle separation. We focus on the experimental realization in Rydberg gases with the associated dipole-dipole interaction $V_{ij}\propto r_{ij}^{-3}$. We impose a lower bound on the distance between pairs of atoms, occurring naturally in experiments with Rydberg atoms \cite{Comparat2010, Signoles2021}, which leads to a particular type of tunable disorder. In this situation all eigenstates are expected to be localized at any disorder strength \cite{Kutlin2020}. However, the extent of eigenstates becomes extremely large at weak disorder \cite{Deng2017}. By extensive numerical simulations using exact diagonalization we study the dependence of dynamical excitation spreading as well as spectral and eigenstate properties of this model as a function of disorder strength and system size. Our main theoretical contribution is to confirm that for this particular type of disorder and geometry all states are localized in the thermodynamic limit for any disorder strength and to provide a microscopic understanding of the eigenstate properties at strong disorder. With regard to an experimental realization of power-law hopping models with Rydberg atom quantum simulators we identify a regime in which localization effects will be observable and highlight potential challenges, thus providing guidance to experimentalists.

The problem of localization in systems with power-law hopping $(\propto r^{-a})$ has been studied theoretically \cite{Yeung1987,Levitov1990,Mirlin1996,Rodriguez2003,deMoura2005,Aleiner2011,Deng2018,Syzranov2019,Kutlin2020,Deng2020} finding localization of all states for the case of $a=3$ and $d=2$ relevant to our study, where $a$ is the power-law exponent of the hopping and $d$ the system dimension. In this case the eigenstates show power-law tails, which means that they are not Anderson localized in the strict sense, which is signaled by exponential localization of the eigenstates. It was found that on-site disorder, as in the case of the Anderson model \cite{Lee1981}, leads to full localization in two dimensions \cite{Eilmes1998, Andrzej2004, Xiong2007}, and to a transition between localized and extended states with a mobility edge in three dimensions \cite{Weaire1977, Economou1977}. These studies considered lattice models where disorder is introduced by imposing uncorrelated random fluctuations on the hopping strengths. A more experimentally realistic scenario is the random placement of atoms with power-law hopping strength between them (power-law Euclidean model). This model has been studied more recently, motivated by experimental advances in quantum simulation with cold molecules and Rydberg atoms \cite{Deng2018, Deng2017, Robicheaux2014, Xiang2013, Yu2016, Botzung2019, Cantin2018, Xu_2015}. 
These works focus on particles in lattice geometries with dilute filling where the disorder strength can be tuned through the filling fraction of the lattice. Rydberg spins randomly placed in continuous space but subject to the Rydberg blockade condition feature a different type of tunable disorder which is less well studied. We are aware of only a study of the spectral statistics \cite{Scholak2014} and recent works studying the appearance of delocalized states, and the role of internal degeneracies for this type of disorder \cite{Abumwis2019, Abumwis2019b, Abumwis2021}. Our work provides a systematic study of the dynamical and eigenstate properties of this model and addresses effects of experimental imperfections under realistic conditions. We confirm that findings of \cite{Deng2017} hold qualitatively also for this kind of disorder; however, the detailed spectral features differ from the lattice case.
First steps towards a realization of this model with Rydberg spins have been reported recently \cite{Guenter2013, Whitlock2019}.

The remainder of our work is structured as follows. In Sec.~\ref{sec:model} we introduce the model and the considered geometry and type of disorder. We study the spectral and eigenstate properties of this model in Sec.~\ref{sec: spectral properties}. We start with a discussion of the dependence on energy and disorder strength and interpret the features observed at strong disorder in terms of small isolated clusters (Sec.~\ref{sec: DOS and IPR}). In Sec.~\ref{sec: Shape} we investigate the localized eigenstates' spatial shape, followed by a study of system-size scaling in Sec.~\ref{sec: Fractality}. Section~\ref{sec: Observing loc effects} is dedicated to the dynamical spreading of an initially localized excitation (Sec.~\ref{sec:unitary_dyn}) and the question of feasibility of experimentally observing localization effects in Rydberg systems (Sec.~\ref{sec:exp_imperfections}). In Sec.~\ref{sec:conclusion} we discuss our results and formulate goals for future research. In the Appendices we back up the results of Sec.~\ref{sec: spectral properties} by providing results on level statistics (Appendix~\ref{sec: LSR}) and by analyzing the properties of low-energy states in detail (Appendix~\ref{sec: Characteristics}).

\section{Model} 
\label{sec:model}

The choice of the model studied in this work is motivated by recent Rydberg atom experiments \cite{Guenter2013, Orioli2018, Whitlock2019, Signoles2021}. The specific setup we consider is a thermal cloud of laser-cooled atoms which, in a first step, are laser excited to a Rydberg state. The created Rydberg atoms are coupled to a second nearby Rydberg state via microwave radiation. The resulting Rydberg spins feature strong dipolar exchange interactions. We will be concerned with the transport of spin excitations within the Rydberg manifold. Atoms that have not been excited to Rydberg states in the initial excitation step are not included in the description. Also, we neglect the thermal motion of the atoms and regard the atomic positions to be fixed, which is a reasonable assumption for typical cloud temperatures and timescales \cite{Signoles2021}. We restrict to the case of a single spin excitation in a two-dimensional geometry. Under these assumptions the system is described by a single-particle hopping model with disorder in the hopping rates. In the following we outline the details of this model and its numerical implementation. 

\begin{figure}
    \centering
    \vspace{-5pt}%
    \hspace{-10pt}%
    \includegraphics{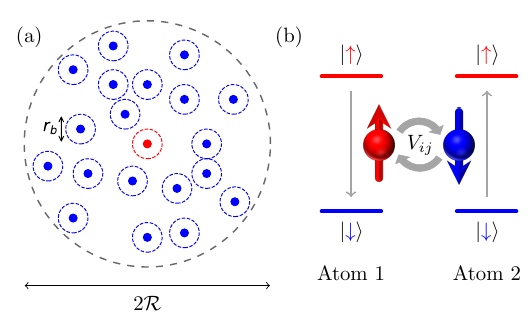}\label{fig:Rydberg Blockade}
\caption{(a) Schematic representation of a 2D cloud of Rydberg atoms respecting the Rydberg blockade constraint and arranged within a spherical volume of radius $\mathcal{R}$. (b) Energy levels of the Rydberg spin system and schematic illustration of dipolar exchange interactions between two Rydberg spins. }\label{fig: Rydberg system}
\end{figure}

\subsection{System geometry}
To model the process of creating Rydberg atoms we randomly place $N$ spins uniformly in a two-dimensional disk-shaped volume of radius $\mathcal{R}$ \footnote{More experimentally realistic geometries and spin density distributions are discussed in Sec.~\ref{sec:exp_imperfections}.}. The excitation of atoms to Rydberg states is subject to the Rydberg blockade constraint \cite{Comparat2010}. Due to the van der Waals interactions between the Rydberg atoms any pair of spins must have a distance larger than the blockade radius $r_b$, which depends on the chosen Rydberg state and the details of the excitation process \cite{Signoles2021}. To model this we draw random spin positions sequentially and reject a sample if its distance to any of the previously drawn positions is less than $r_b$. This procedure is equivalent to randomly placing $N$ disks of diameter $r_b$ in a given volume, and is known as random sequential absorption \cite{Hinrichsen1990, Cadilhe2007}. A typical positional configuration generated in this way is shown in Fig.~\ref{fig: Rydberg system}\hyperref[fig: Rydberg system]{(a)}.

Due to the blockade constraint atom positions are not completely random, resulting in the degree of disorder being tunable. While at low number density, i.e., when the ratio of blockade radius over mean interparticle spacing is small, atom positions are uncorrelated (strong disorder), at higher density spin positions become more densely packed and regular structures appear (weak disorder). The densest packing of disks in two dimensions is realized for a regular hexagonal lattice configuration. However, the random sequential absorption process reaches the so-called jamming limit at which no further atom can be placed. Defining the filling fraction (dimensionless density) $\rho=N(r_b/2)^2/\mathcal{R}^2$ as the ratio between the area covered by the disks and total area, the jamming limit is given by $\rho_{\rm max}= 0.5472 \pm 0.0002$ \cite{Hinrichsen1990}. The number of random trials necessary for generating samples of density $\rho$ increases as $(\rho_{\rm max}-\rho)^{-2}$ \cite{Cadilhe2007} as one approaches the jamming limit. In the interest of keeping computing time reasonable we investigate densities up to $\rho = 0.53$. Experimentally, the density can be tuned by varying the strength and duration of the laser pulses that are used to excite the atoms to Rydberg states, as well as the density of ground state atoms which puts an upper bound on the reachable Rydberg atom density.

\subsection{Hamiltonian}

The excitation transport takes place in the pseudo-spin-1/2 system where the spins are encoded in two Rydberg states
\begin{align}
\ket{\downarrow}= \ket{nS_\frac{1}{2},m_j=+1/2} \,, && \ket{\uparrow}= \ket{nP_\frac{3}{2},m_j=+3/2} \,.
\end{align} 
In the initial excitation step, described in the previous subsection, Rydberg atoms are created in the state $\ket{\downarrow}$. Subsequently, spin excitations can be created by microwave coupling between the two spin states \cite{Orioli2018, Whitlock2019, Signoles2021}. The dynamics ensuing from dipolar exchange interactions between the spin states [see Fig.~\ref{fig: Rydberg system}\hyperref[fig: Rydberg system]{(b)}] is described by the XY spin Hamiltonian
\begin{align}
\label{eq:Hamiltonian}
H=-\frac{1}{2}\sum_{i\neq j}^N V_{ij}(S_i^+S_j^-+S_i^-S_j^+) \,,
\end{align}
where $S^\pm_i=S^x_i\pm iS^y_i$ are the spin raising and lowering operators corresponding to atom $i$ with $S^\alpha_i$ $(\alpha=\{x,y\})$ being the spin-1/2 angular momentum operators.
$V_{ij}$ is the matrix element of the dipole-dipole interaction
\begin{align}
\label{eq:Coupling}
V_{ij}=\hbar C_3 \frac{ (1-3\cos^2{\theta_{ij}})}{|\mathbf{r}_i-\mathbf{r}_j|^3} \,,
\end{align}
where $\theta_{ij}$ denotes the angle between the quantization axis and the difference vector $(\mathbf{r}_i-\mathbf{r}_j)$ between the atom positions. In our two-dimensional geometry we choose the quantization axis to be perpendicular to the plane ($\theta=\frac{\pi}{2}$) leading to isotropic power-law interactions $V_{ij}=\hbar C_3/|\mathbf{r}_i-\mathbf{r}_j|^3$.

The dynamics under the spin Hamiltonian \eqref{eq:Hamiltonian} conserves the number of excitation, i.e., the number of spins in the state $\ket{\uparrow}$. We restrict to the case of a single excitation, in which the problem takes the form of a hopping model:
\begin{align}
\label{eq:HoppingHam}
H=-\sum_{i\neq j}^N V_{ij} \ket{i}\bra{j} \,,
\end{align}
where $\ket{i}$ describes the state in which atom $i$ is in state $\ket{\uparrow}$, while all other atoms are in state $\ket{\downarrow}$. The spin excitation takes the role of a particle hopping between the sites of a random graph given by the interaction strengths $V_{ij}$.
We note that in this model the Hamiltonian is composed of off-diagonal terms only, in contrast to the standard Anderson hopping model where nearest neighbor hopping is combined with disorder generated by randomizing the on-site potentials. 

Unless specified otherwise, we will use dimensionless units by setting $r_b$, $C_3$, and $\hbar$ to unity, which sets $C_3/(\hbar r_b^3)$ to be the unit of energy and $r_b^3/C_3$ to be the unit of time. Typical experimental values are $C_3/2\pi=0.86$ GHz $\mu \mathrm{m}^3$ and $r_b = 5 \; \mu \mathrm{m}$ (using $n=48$) \cite{Orioli2018}, which is, however, largely tunable by choosing Rydberg states with different principal quantum number. After rescaling to these units, the dimensionless density $\rho$ and the number of atoms, or sites, $N$ remain as free model parameters. 
In the following sections we investigate the impact of off-diagonal disorder in the hopping terms $V_{ij}$ generated by the random atom positions on excitation transport for varying density, i.e., disorder strength, and system size. For this we numerically solve the hopping model \eqref{eq:HoppingHam} by exact diagonalization of the matrix $V_{ij}$ for system sizes up to $N = 32000$ atoms.

\section{Spectral and eigenstate properties}
\label{sec: spectral properties}

\begin{figure*}
    \centering
\includegraphics[scale=0.5]{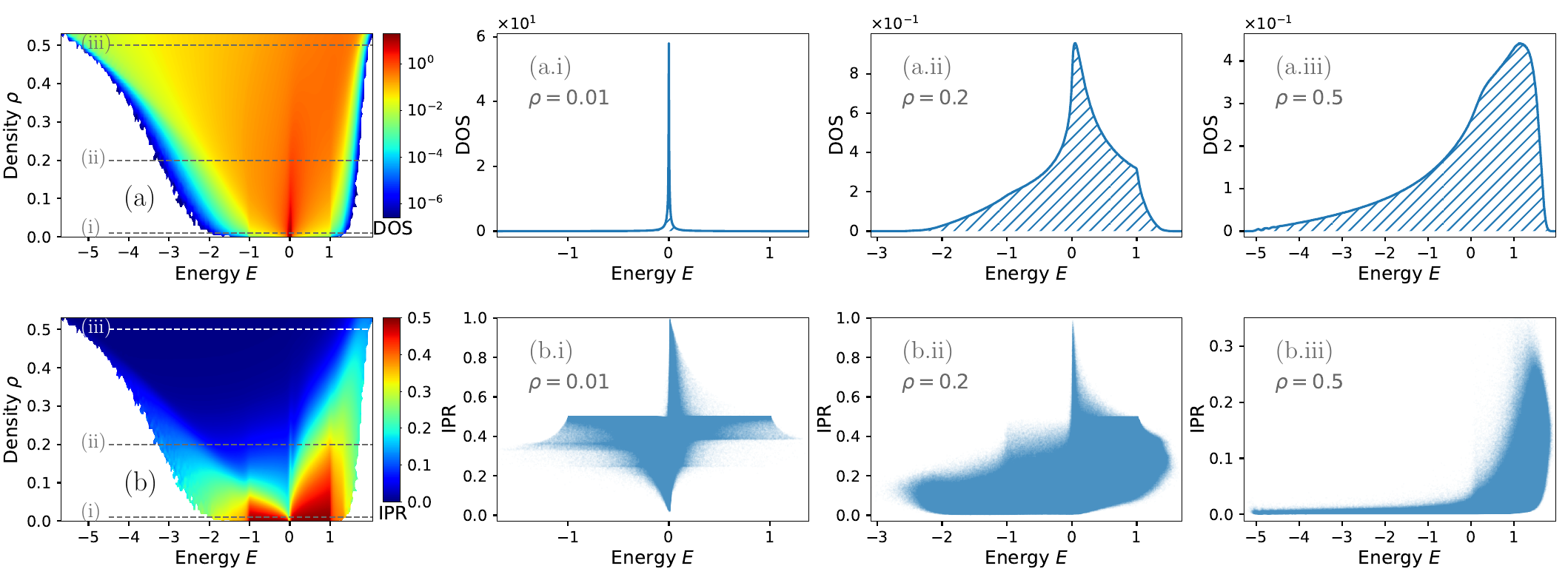}
\caption{Overview of the density dependence of spectral and eigenstates properties. (a) DOS for all densities. (a.i), (a.ii), (a.iii) Cuts at densities $\rho=0.01,\; 0.2$ and  $0.5$, respectively.
(b) Energy-binned IPR for all densities. (b.i), (b.ii), (b.iii) Eigenstate IPRs at densities $\rho=0.01,\; 0.2$ and  $0.5$. We used $N= 2000$ atoms, adjusted  the  system  size $\mathcal{R}$ to match each density $\rho$ and averaged over $50000$ disorder realizations. The energy was divided into $200$ bins which amounts to $\Delta E \approx 0.039$ for (a) and (b) and $1000$ bins with $\Delta E \approx 0.0034$ for (a.i)--(a.iii). For (b.i)--(b.iii) we included $12000$ different disorder realizations.}\label{fig: Eigenstates}
\vspace{-1mm}
\end{figure*}

In this section we study the properties of the eigenstates of the Euclidean hopping model Eq.~\eqref{eq:HoppingHam} focusing on localization effects. For the Anderson model in $d$ dimensions with power-law hopping $V\propto r^{-a}$ it is known that for $a>d$ and $d\leq2$ all states are localized for any disorder strength. Interestingly, for $a<3d/2$ a set of extended states exists \cite{Rodriguez2003,deMoura2005}. The fraction of these states, however, scales sublinearly with the system size and is thus expected to vanish in the thermodynamic limit. Recently, a renormalization method has been develop \cite{Kutlin2020} showing the equivalence between translation invariant models with diagonal disorder and Euclidean models. This work indicates that Euclidean models in $d\leq2$ dimensions feature at most a set of measure zero of delocalized states in the thermodynamic limit. Numerical experiments showed that, indeed, all states are localized \cite{Deng2017}, for the present case of $d=2$ and $a=3$. However, in Ref.~\cite{Deng2017} a specific kind of disorder was used, and it was found that localization lengths can be extremely large in 2D leading to an effective localization-delocalization crossover at realistic system sizes. 

To address the question, whether the same occurs for the type of disorder present in Rydberg systems, we study spectral and eigenstate properties systematically, as a function of density $\rho$ (i.e., disorder strength), energy, and system size. 
Our results overall confirm that the findings of Ref.~\cite{Deng2017} also apply to the blockade geometry -- all states are localized in the thermodynamic limit but a localization to delocalization transition as a function of density occurs at finite system size -- and complements previous work by providing a detailed analysis of the spectral and eigenstate properties. In particular, we undertake a detailed study of the spectral features at low densities in terms of small clusters and discuss the spatial shapes of eigenstates at high and low densities.
We find that at low densities almost all states are localized in the sense that their spatial extents are much smaller than the system size and generally decay spatially with a power law, with exceptions in the middle of the spectrum. For these systems all spectral features can be understood in terms of small clusters of strongly interacting spins. At high densities the eigenstates in the bulk of the spectrum, i.e., around the maximum of the density of states (DOS), are still localized within small regions, while at low energies a large tail of seemingly extended states develops. However, a careful analysis of the system-size dependence of the eigenstate participation ratio indicates that the finite-size generalized fractal dimension decays in the large $N$ limit throughout the spectrum, which suggests that all states will eventually become localized in the infinite-size limit.

\vspace{-1mm}
\subsection{Density of states and inverse participation ratio}
\label{sec: DOS and IPR}

The main property of interest for studying localization is the inverse participation ratio (IPR) of eigenstates. Before presenting our numerical results we briefly introduce this quantity. 
The IPR of a state $\ket{\psi}=\sum_j c_j\ket{j}$ is defined as $\mathrm{IPR}=\sum_j |c_j|^4$. Its inverse, the participation ratio (PR), quantifies how many basis states participate in the state, or in the language of particle transport, over how many sites the particle is distributed. For a state perfectly localized on site $k$, i.e., $c_j=\delta_{jk}$, one has $\mathrm{PR}=1$, while for a state completely delocalized over all $N$ sites, $c_j=1/\sqrt{N}$, we obtain $\mathrm{PR}=N$. 
Accordingly, the IPR can take values $1/N \leq$ IPR $\leq 1$ and is large for localized states and small for extended ones. 
In the following we use the IPR to investigate whether eigenstates $\ket{\phi_n}=\sum_j c_j^{(n)}\ket{j}$ at eigenenergies $E_n$ are of localized or extended nature and how their properties depend on the atom density $\rho$, i.e., on the disorder strength. We note that usually, the terms localized and extended, refer to the system-size scaling of eigenstate IPRs. These properties will be discussed in Sec.~\ref{sec: Fractality}, while in the present subsection we will refer to localized (extended) states as states with $\mathrm{IPR}\sim 1$ ($\mathrm{IPR}\sim 1/N$), respectively.  We also note that in the context of Anderson localization only states with exponentially decaying probability density are called localized. Here, instead we can expect only algebraic localization.

We begin our numerical study by examining the density of states, $\mathrm{DOS}(E)=|K_E|$ with $K_E = \{n \; \big| \; E-\Delta E/2\leq E_n < E+\Delta E/2 \}$, as a function of energy and atom density. In Fig.~\ref{fig: Eigenstates}\hyperref[fig: Eigenstates]{(a)} we used $N=2000$ atoms, segmented the energy into 200 bins (corresponding to a bin size $\Delta E\approx0.039$), adjusted the system size $\mathcal{R}$ to match each density $\rho$, and averaged over $50000$ disorder realizations, i.e., random atom placements. Figures~\ref{fig: Eigenstates}\hyperref[fig: Eigenstates]{(a.i)--}\ref{fig: Eigenstates}\hyperref[fig: Eigenstates]{(a.iii)} show cuts at three different densities as indicated by the dashed lines in Fig.~\ref{fig: Eigenstates}\hyperref[fig: Eigenstates]{(a)}.
The DOS is symmetric at low atom density, i.e., strong disorder, and sharply peaks at $E=0$. It broadens and becomes asymmetric at higher densities, with a long tail at negative energies.

These features can be understood microscopically in terms of small clusters of regularly spaced atoms. At low density the probability to encounter clusters of multiple atoms forming a regular structure is low. Thus the spectrum is dominated by dimers, i.e., pairs of closely spaced atoms $i$ and $j$, with the atoms surrounding the dimer much further away from it than the spacing of the pair. In this case we can treat the interactions of the dimer with the remainder of the system as a perturbation to the dimer Hamiltonian $H_{ij}=-V_{ij}(\ket{i}\bra{j}+\ket{j}\bra{i})$. Neglecting the interaction of the dimer with its surrounding completely results in the eigenstates $\ket{\phi_{\pm}}=(\ket{i}\pm\ket{j})/\sqrt{2}$ being perfectly localized on the dimer and eigenenergies $E_{\pm}=\mp V_{ij}$ symmetrically distributed around $E=0$. This explains the symmetric shape of the DOS at low densities. We note that the dimer picture has also been employed in previous works to explain spectral properties of power-law hopping models \cite{Scholak2014, Abumwis2019, Abumwis2019b}. In Fig.~\ref{fig: Eigenstates}\hyperref[fig: Eigenstates]{(a)} we also observe that the DOS becomes narrower and more strongly peaked at $E=0$ the lower the density, which is explicitly shown in Fig.~\ref{fig: Eigenstates}\hyperref[fig: Eigenstates]{(a.i)} for $\rho=0.01$. This is due to the fact that the average pair distance is simply larger for lower densities and thus the average interaction energy becomes smaller. In fact, for very low densities the blockade effect becomes negligible resulting in uncorrelated atom positions. In this limit, the problem becomes scale invariant as the Hamiltonians for different densities are equal up to a global rescaling of energy. Thus, further reducing the density results in a narrowing of the DOS while leaving its shape unchanged.

\begin{figure}
\vspace{0.5mm}
\includegraphics[trim =1pt 0 0 0]{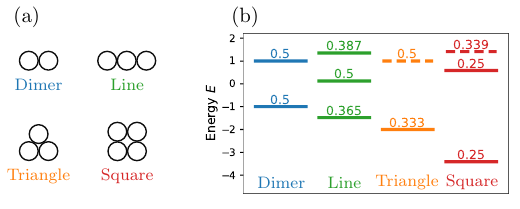}
\caption{Examples of small regular clusters.
(a) Different types of regular arrangements, where the smallest distances is given by $r_b$. (b) The corresponding eigenenergies. The numbers above the lines denote the IPR for the respective state. Dashed lines indicate doubly degenerate levels. The ground state has equal occupations and sign on all sites except for the case of the line, where the central site has a higher occupation, thus not reaching the minimal IPR of $1/N$.  }\label{fig:Micro Ansicht}
\end{figure}

The highest dimer energy is realized if the pair distance equals the lower distance cutoff given by the Rydberg blockade $r_b$, which gives $V_{r_b}=C_3/r_b^3=1$ in our units. This constraint explains the drop of the DOS for $|E|>1$ visible in Figs.~\ref{fig: Eigenstates}\hyperref[fig: Eigenstates]{(a)} and \ref{fig: Eigenstates}\hyperref[fig: Eigenstates]{(a.ii)}. Energies outside of this window can be attained only by larger clusters of atoms separated by distances close to the blockade radius. In Fig.~\ref{fig:Micro Ansicht} we show the eigenenergies of a selection of such clusters. This shows that for growing cluster size the eigenenergies extend much further towards negative energies than towards positive ones. The observed asymmetry is an intrinsic property of the purely negative and off-diagonal hopping Hamiltonian \eqref{eq:HoppingHam}. At higher densities the eigenstates become increasingly delocalized over larger clusters of atoms explaining the observed overall asymmetry in the DOS. Also, the peak of the DOS around $E=V_{r_b}=1$ observed at high densities [Fig.~\ref{fig: Eigenstates}\hyperref[fig: Eigenstates]{(a.iii)}] has a precursor in the spectra of small clusters as their eigenstates show degeneracies at this energy visible in Fig.~\ref{fig:Micro Ansicht}.

We now turn to the eigenstate IPR shown in Fig.~\ref{fig: Eigenstates}\hyperref[fig: Eigenstates]{(b)}. At low densities, corresponding to uncorrelated atom positions and thus strong disorder, all states are localized, except for a few states close to $E=0$ with small IPR. Since in power-law hopping models with $a<3d/2$ a set (of measure zero of) of truly delocalized states exist \cite{Rodriguez2003,deMoura2005,Kutlin2020}, it is plausible that for our case of $a=3d/2$ there are states with small IPR which appear delocalized for any finite system size even at the lowest densities. For higher density, i.e., increasingly dense packing of atoms, the IPR globally decreases indicating that eigenstates become more extended. While in the spectral bulk at positive energies most states are still fairly localized, a strong tail of delocalized states emerges at negative energies [see Fig.~\ref{fig: Eigenstates}\hyperref[fig: Eigenstates]{(b.iii)}]. At low densities, the energy-binned IPR in Fig.~\ref{fig: Eigenstates}\hyperref[fig: Eigenstates]{(b)} shows a sharp feature at $|E|=1$, the blockade energy, which can be understood in terms of the cluster picture developed above. In Figs.~\ref{fig: Eigenstates}\hyperref[fig: Eigenstates]{(b.i)--(b.iii)} we show the IPRs of all eigenstates without binning for specific densities, which reveals an even richer structure, which can be fully understood in terms of our microscopic picture, as discussed in the remainder of this section. 

In Fig.~\ref{fig: Eigenstates}\hyperref[fig: Eigenstates]{(b.i)} we observe a sharp horizontal feature for $|E|<1$ where states accumulate at $\mathrm{IPR}\lesssim 1/2$. This corresponds to the IPR of dimers. The better they are isolated from their surrounding the closer the IPR of states localized on them is of the maximal value of $1/2$. Dimer energies are confined to $|E|\leq 1$ due to the blockade constraint, which explains that states at even higher energies must be due to larger clusters and have smaller IPR. Indeed, we observe accumulations of points at $\mathrm{IPR}\approx 1/3$ and $1/4$ due to trimers and tetramers, which extend further towards negative energies. At $E>0$ we observe a feature at $\mathrm{IPR}\approx 0.387$, which is the IPR corresponding to the highest energy state of a trimer, realized by the line configuration as shown in Fig.~\ref{fig:Micro Ansicht}. Around $E=0$ we observe a substantial number of more strongly delocalized states with $\mathrm{IPR}\ll 1/2$. Their distribution in energy again shows an asymmetry with a tail towards negative energies as expected from the analysis of the cluster spectra in Fig.~\ref{fig:Micro Ansicht}. 

Interestingly, around $E=0$ we observe a group of states that is even more localized than the dimer eigenstates, i.e., $\mathrm{IPR}>1/2$. These states result from atoms being separated further from their surrounding atoms than the typical distance among those. In Fig.~\ref{fig: Eigenstates}\hyperref[fig: Eigenstates]{(b.i)} we observe that these states appear exclusively at positive energies, which can be understood in a perturbative picture: We consider an isolated atom $k$ being weakly coupled to a small cluster, exemplified here by a dimer of atoms $i$ and $j$. Written in the eigenbasis of the dimer Hamiltonian, using that $V_{ik}\approx V_{jk}\equiv V$, the Hamiltonian of this system reads
\begin{equation}
\label{eq:dimer_pert}
\begin{split}
    H = -V_{ij} & (\ket{\phi_{+}}\bra{\phi_{+}} -\ket{\phi_{-}}\bra{\phi_{-}} ) \\
    & - (V\sqrt{2}\ket{\phi_{+}}\bra{k} + \mathrm{H.c.}) \,.
\end{split}
\end{equation}
This shows that the state $\ket{k}$ couples to only the, energetically lower, symmetric state of the dimer. Thus the level repulsion due to the coupling between $\ket{\phi_+}$ and $\ket{k}$ will lead to an upward shift of the state localized on the single site $k$. If $V_{ik}$ and $V_{jk}$ are only approximately equal, the coupling of $\ket{k}$ to $\ket{\phi_+}$ will still be much stronger than to $\ket{\phi_-}$ shifting $\ket{k}$ to positive energies. This argument can also be extended to the perturbative coupling of a single site to larger clusters. 

We note that for the case of a partially filled lattice geometry studied in Ref.~\cite{Deng2017} the spectral features at low densities will differ substantially due to the geometric constraints on the small clusters imposed by the lattice geometry.

At intermediate densities, shown in Fig.~\ref{fig: Eigenstates}\hyperref[fig: Eigenstates]{(b.ii)}  most of the features caused by small clusters are still visible, but we now observe an increasing fraction of states with very small IPR at $E<0$. These delocalized states become even more prominent at the highest densities, while states at $E>0$ still stay rather localized; see Fig.~\ref{fig: Eigenstates}\hyperref[fig: Eigenstates]{(b.iii)}. One might suspect that a crossover from a localized to a partly delocalized phase with a mobility edge at $E=0$ occurs as density is increased. However, as we show in Sec.~\ref{sec: Fractality} a detailed analysis of the system-size dependence of the IPR suggests that all states will become localized eventually, even for the highest densities.

\subsection{Spatial shape of eigenstates}
\label{sec: Shape}

Before proceeding to the analysis of the system-size dependence of the IPR, we briefly discuss the spatial shape of the localized eigenstates. We define the radial density $n(r)$ of a state as the average excitation probability of an atom at distance $r$ from the position $\mathbf r_m$ of the state's highest occupied site, i.e., $|c_m^{(n)}|^2$ is largest among the populations $|c_j^{(n)}|^2$. Defining the set of indices $K_r = \{j \; \big| \;  |\mathbf{r}_j-\mathbf{r}_{m}| \in [r,r+\delta r)\}$ of atoms inside an annulus around $\mathbf{r}_{\rm m}$, we write the radial density of an eigenstate $\ket{\phi_n}$ as
\begin{align}
\label{eq: radial density}
n(r,\phi_n) = \frac{1}{|K_r|} \sum_{j \in K_r} |c_j^{(n)}|^2 \,,
\end{align}
where $|K_r|$ is the number of elements in the set.

\begin{figure}
    \centering
\includegraphics[scale=0.5]{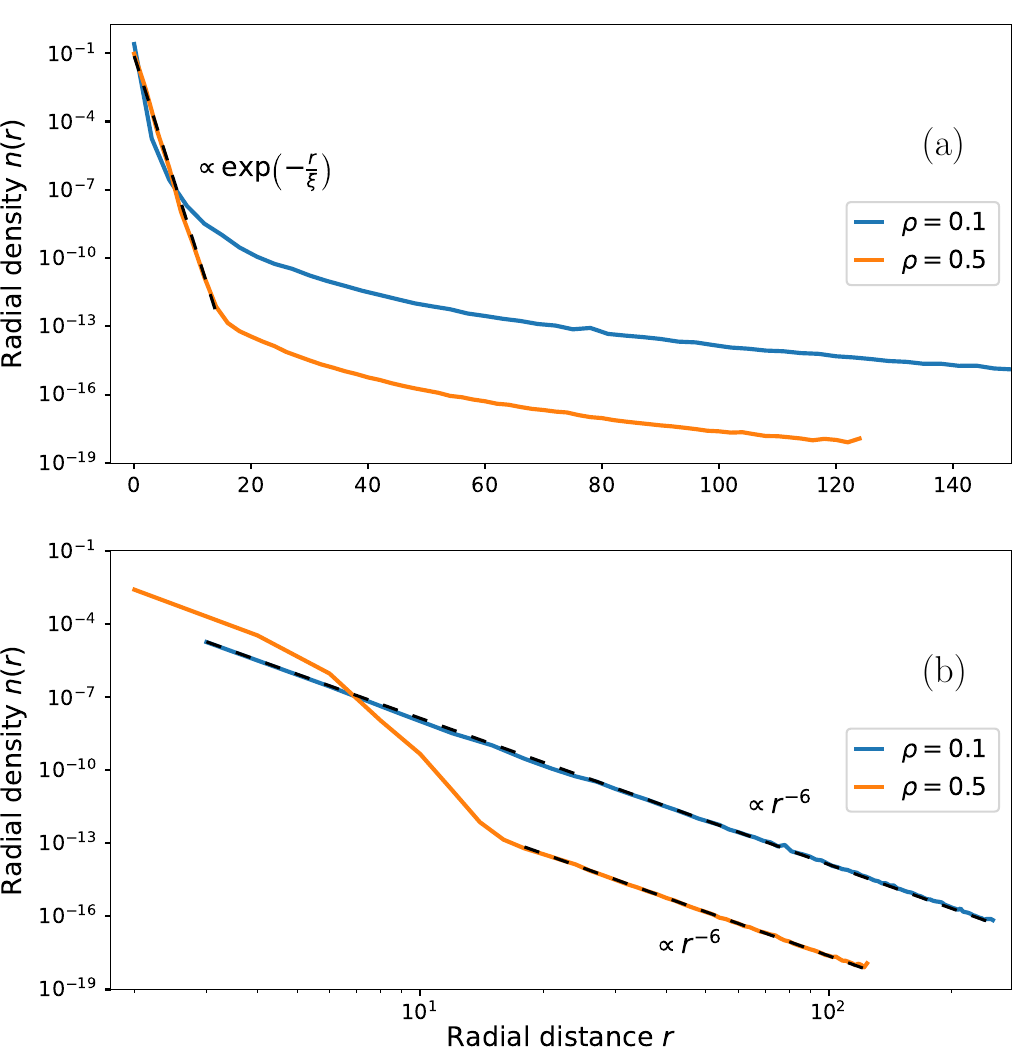}
\caption{Radial density of highest energy eigenstates for $N=10000$  at $\rho=0.1$ and $\rho=0.5$ in a single logarithmic (a) and in a double logarithmic (b) plot. Note that here only a single state was chosen, hence no disorder averaging was performed. We chose the radial bin size to be $\delta r=3$ for the case $\rho=0.1$ and $\delta r=2$ for $\rho=0.5$. While for the small density (blue) the state decays entirely with power law $r^{-6}$, the state for the high density (orange) shows an exponential onset [see dashed line in (a)] with a subsequent power-law tail.
\vspace{-3mm}
}
\label{fig:Shape Eigenstates}
\end{figure}

We find that strongly localized eigenstates ($\mathrm{PR}\ll N$) typically have power-law tails. This characteristic feature of power-law Euclidian models \cite{Yeung1987,Kutlin2020,Modak2021} is in contrast to Anderson localization, where states are exponentially localized allowing to define a localization length. In Fig.~\ref{fig:Shape Eigenstates} we show the radial density for the highest energy state at low and high density as an example. In the double logarithmic plot [Fig.~\ref{fig:Shape Eigenstates}\hyperref[fig:Shape Eigenstates]{(b)}], the $r^{-6}$ power-law tails are clearly visible.
Interestingly, for the state at high density we find an initial exponential decay \cite{Yeung1987,Modak2021}, as seen in the single-logarithmic plot [Fig.~\ref{fig:Shape Eigenstates}\hyperref[fig:Shape Eigenstates]{(a)}]. Such an exponential onset was observed for all the localized states at $E>0$ for high densities, however, it becomes less pronounced as $E=0$ is approached from above. The localization length $\xi$ varies between eigenstates and correlates with the eigenstates' PR. The extent of the exponential head of the highest excited state grows with increasing system size, indicating that in the thermodynamic limit true exponential (Anderson) localization is recovered. At low densities only some of the least localized states around $E=0$ show this behavior. This emergent exponential localization also leads to the surprising observation that the long-distance tail of the radial density for the high-density case falls below that of the low-density case. Understanding the dependence of the emerging localization length scale $\xi$ on energy and density and its microscopic origin requires further investigation.  

Power-law tails with $r^{-6}$ decay are found not only for the highest excited state but, at low density, also for all eigenstates at $E<0$ with an IPR close to $0.5$, i.e., the $\ket{\phi_+}$ states of dimers. For dimer states at $E>0$ we mostly find power-law tails with larger exponents, but for less localized states also power-law tails with smaller exponents occur, and even nonmonotonous behavior is encountered.
Algebraic tails $\propto r^{-2a}$ emerging in power-law hopping models have been observed for $d=1$ \cite{Yeung1987, Deng2018} and can be understood within a perturbative picture for $a>d$.
For the example of dimer states, this perturbative approach implies, according to Eq.~\eqref{eq:dimer_pert}, that predominantly the symmetric eigenstate $\ket{\phi_+}$ (with eigenenergy $-V_{ij}$) couples to a distant third atom, explaining the observation that only dimers at $E<0$ show clean $r^{-6}$ tails.
We note that the highest excited state shown in  Fig.~\ref{fig:Shape Eigenstates} for the low-density case is actually a trimer in the "line" configuration, for which the highest energy state does couple to distant single sites.

\begin{figure*}
    \centering
{\includegraphics[scale=0.5]{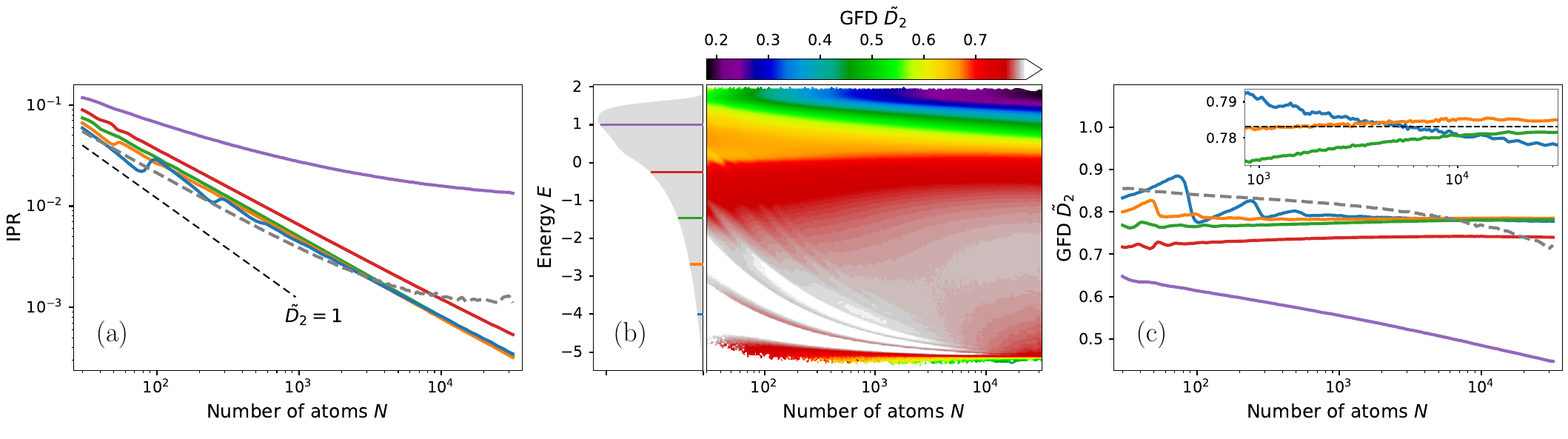}}
    \caption{System-size dependence of IPR and GFD for $\rho=0.5$. (a) IPR at different energies [color encoded; see panel (b)]. The dashed gray line shows the behavior of the ground state. (b) Density plot of the energy-binned GFD $\tilde{D}_2$. The panel on the left shows the DOS for $N=32000$ with the energies corresponding to the courses of IPR from (a) highlighted as colored lines. (c) Generalized fractal dimension for selected energy windows matching those shown in (a). The inset shows an enlargement of the data around the apparent asymptotic value. The horizontal dashed line serves as a reference. We used atom numbers between $N=30$ and $N=32000$, and the number of disorder realizations interpolates between $10^6$ for $N=30$ and $200$ for $N=32000$. The energy was divided into 200 bins, which amounts to $\Delta E \approx 0.038$.
    }
    \label{fig: Frac}
\end{figure*}

\subsection{Finite-size scaling analysis}
\label{sec: Fractality}

We now turn to the question whether the delocalized states observed at high densities are truly extended in the sense that their extent scales with the system size.
We  characterize the nature of an eigenstate $\ket{\phi_n} = \sum_j c_j^{(n)} \ket{j}$ by considering the asymptotic scaling of the moments
\begin{align}
\label{eq: Fractality}
 I_q(\phi_n) = \sum_j |c_j^{(n)}|^{2q} \propto N^{-\tau_q(\phi_n)} \,, \qquad q\in \mathbb{R^+}
\end{align}
with system size $N$ \cite{De_Luca_2014, Wegner1981}. For $q=2$ we recover the inverse participation ratio $I_2=\mathrm{IPR}$ defined above.
For localized states $I_q$ is independent of system size and thus $\tau_q = 0$ for any $q \geq 1$. For ergodic states, which are spread out over the entire system, one obtains $\tau_q=q-1$ for all $q$. In particular, the IPR scales as $N^{-1}$ as discussed above. The asymptotic scaling behavior of $I_q$ is conveniently described by the fractal dimension $D_q= \tau_q/(q-1)$, such that $D_q=1$ for ergodic states and $D_q=0$ for localized states. Nonergodic extended states, with $0<D_q<1$ being $q$-dependent, are called multifractal. In order to assess the asymptotic ($N \rightarrow \infty$) scaling based on exact diagonalization data at finite $N$, we introduce the finite-size generalized fractal dimensions (GFD) \cite{Lindinger2019, Rodriguez2011, Pausch2020}
\begin{align}
 \tilde{D}_q(\phi_n)= \frac{1}{1-q} \log_N I_q(\phi_n)
\end{align}
Since $D_q=\lim_{N \rightarrow \infty} \tilde{D}_q$ one can uncover multifractal behavior if $\tilde{D}_q$ saturates at some finite value in the limit of large $N$. 
In the following we investigate this scaling behavior restricting to the case of $q=2$, i.e., the IPR, and focusing on the highest considered density of $\rho=0.5$.

The system-size dependence of the IPR is shown in Fig.~\ref{fig: Frac}\hyperref[fig: Frac]{(a)} averaged over states with eigenenergies in a window around selected energies [color encoded as indicated in Fig.~\ref{fig: Frac}\hyperref[fig: Frac]{(b)}]. In the double-logarithmic plot multifractal behavior would be visible as an asymptotically linear dependence with a slope between $0$ (localized) and $-1$ (ergodic, dashed black line). It indeed seems that for states at $E>0$, after an initial decrease, the slope of the IPR, representing the fractal dimension, approaches zero at large $N$ [purple line in Fig.~\ref{fig: Frac}\hyperref[fig: Frac]{(a)}] indicating full localization, while at $E<0$ states with finite fractal dimension appear. In turn, the ground state (dashed gray line), after an initial linear decrease, clearly becomes localized. 

To further scrutinize these observations we show the GFD $\tilde{D}_2$ as a function of energy and system size in Figs.~\ref{fig: Frac}\hyperref[fig: Frac]{(b)} and \ref{fig: Frac}\hyperref[fig: Frac]{(c)}.
Close inspection of the $N$-dependence reveals that the seemingly multifractal states in the low-energy tail of the spectrum do not converge to a constant value of $\tilde{D}_2$ at large $N$. In particular, we find a decreasing trend for states at $E<0$ for $N>10^3$ [red line in Fig.~\ref{fig: Frac}\hyperref[fig: Frac]{(c)}] after an initial increase. Thus, the apparent mobility edge turns out to soften at large $N$ indicating that $E=0$ ceases to be a special point asymptotically. Also, at very small energies (blue line) the GFD decreases at large $N$, visible also in Fig.~\ref{fig: Frac}\hyperref[fig: Frac]{(b)} as a region of small GFD at low energy which grows with $N$. The ground state GFD shows a globally decreasing trend. Around $E=-2$ we observe a spectral region where the GFD is still increasing with $N$ up to the largest system sizes considered. However, the inset of Fig.~\ref{fig: Frac}\hyperref[fig: Frac]{(c)} shows that the GFD is concave in all cases suggesting an eventual decrease. We have also studied the $N$-dependence of the GFD at a slightly smaller density of $\rho=0.45$ where at large system sizes a downward trend is found at all energies.
We take these numerical observations as evidence that eventually all states become localized even at the highest densities. Given that at $a<3d/2$ a set of truly delocalized states exists at any finite system size \cite{Kutlin2020} it is plausible that in the present case of $a=3d/2$ extremely large system sizes may be required to numerically see localization of all states.

Another indicator of localization is the level spacing statistics. Specifically, the level spacing ratio is expected to take certain values for localized and ergodic systems, respectively, predicted by random matrix theory. In Appendix \ref{sec: LSR} we show that at finite system sizes parts of the spectrum show level statistics close to the expectation for ergodic states. However, the level spacing ratio globally (at all energies) decreases towards the localized value asymptotically at large $N$, confirming our conclusion that all states eventually become localized. 

Finally, we note that the oscillations in $\tilde{D}_2$ visible at low energies and small $N$ in Fig.~\ref{fig: Frac}\hyperref[fig: Frac]{(b)} are due to finite-size effects, as explained in detail in Appendix \ref{sec: Characteristics}. In brief, the lowest lying eigenstates can be understood in terms of a quasicontinuous picture. States minimize their energy by localizing at the minimum of a mean-field potential. For small system sizes, the walls of this mean-field potential are given by the system boundaries, which leads to an energy gap between the ground and excited state within this potential, visible as oscillations in the density of states and IPR. This also explains that, e.g., the blue line in Fig.~\ref{fig: Frac}\hyperref[fig: Frac]{(a)} shows a piecewise ergodic behavior: Each section of linear decrease corresponds to an individual low-lying state, e.g. the ground state, which explores the full system and thus grows with system size.

In conclusion, our numerical results indicate that all eigenstates are localized in the limit $N\rightarrow \infty$, but their extent can be extremely large at high densities. This has implications for experiments, which are naturally limited in system size. Localization effects will be difficult to observe experimentally in this regime as discussed in detail in the following section.

\section{Observing localization effects}
\label{sec: Observing loc effects}

The spectral and eigenstate properties discussed in the previous section cannot be probed directly experimentally. In this section we explore how localization effects manifest in the dynamical spreading of an initially localized excitation.
We find that for low densities and sufficiently large system size localized eigenstates cause excitation spreading to halt before reaching the system boundary. At high densities the extent of the wave function at late-times scales with system size as a result of finite-size delocalized eigenstates. At low densities the late-time excitation density decays radially following a stretched exponential function. Turning to realistic experimental scenarios we find that decoherence effects will strongly limit the regime in which localization effects are observable. Decoherence is found to lead to subdiffusive excitation spreading and eventual complete delocalization. This finding underlines that the localization observed in the unitary case is a true quantum interference effect, which is destroyed when reducing the coherence between different hopping paths.

\subsection{Unitary dynamics}
\label{sec:unitary_dyn}

We start by comparing the propagation of an initially localized excitation for different densities. The initially excited atom is placed in the center of the system for each disorder realization. In the following, we refer to this initial state as $\ket{\psi_0}$. The dipole-dipole interactions will now lead to an expansion and spread of the excitation. For $t \rightarrow \infty$ the expansion can either reach the boundaries of the system or localization effects prevent further spreading. The occurrence of the latter is illustrated in the time evolution shown in Fig.~\ref{fig: Expansion + MSD}\hyperref[fig: Expansion + MSD]{(a)}.

We determine the population of all sites, i.e., the probabilities $P_j(t)=|\bra{j}\exp(-iHt)\ket{\psi_0}|^2$ for finding the excitation on atom $j$ after time $t$, for a discrete grid of times. 
These populations are obtained straight forwardly from the eigenvalues $\{E_n \mid 1 \leq n \leq N\}$ and eigenstates $\{\ket{\phi_n} \mid 1 \leq n \leq N\}$ of the Hamiltonian \eqref{eq:HoppingHam} as 
\begin{equation}
\label{eq: population}
P_{j}(t) = \biggl| \sum_n e^{-iE_nt} \bra{j}\ket{\phi_n} \bra{\phi_n}\ket{\psi_0}\biggr|^2 \,.
\end{equation}
In the following we will compare not only time dependent propagation, but also differences in the asymptotic late-time population, defined via $P_{j}(t\rightarrow \infty) = \lim_{T\rightarrow \infty} \frac{1}{T} \int_0^T dt \; P_{j}(t)$. Using the expression for the population from Eq.~\eqref{eq: population} and exploiting that there are no degeneracies, this results in the diagonal ensemble expectation 
\begin{equation}
 P_{ j}(t\rightarrow \infty) =   \sum_{n} \bigl| \bra{j}\ket{\phi_n}\bigr|^2 \bigl| \bra{\phi_n}\ket{\psi_0}\bigr|^2 \,,
 \label{eq:p_inf}
\end{equation}
which depends on only the properties of the eigenstates overlapping with the initial state.

To further quantify the spreading of the excitation we introduce the mean square displacement (MSD) $\langle r^2(t) \rangle$, which characterizes the mean expansion and is the most common measure of the spatial extent of random motion. The MSD is given by $\langle r^2(t) \rangle = \sum_{j=1}^N r_j^2 P_j(t)$, where $r_j = |\mathbf{r}_j|$ is the distance of atom $j$ to the position of the initial excitation at the origin. The MSD is bounded above due to the finite size of our system. Its maximal value $\langle r^2 \rangle_{\rm max}=\mathcal{R}^2/2$ is obtained for a homogeneous probability distribution with $P_j = 1/N \; \forall j$.

\subsubsection{Temporal excitation spreading}

\begin{figure}
        \centering
       \includegraphics[trim=3.5pt 0 0 0 ]{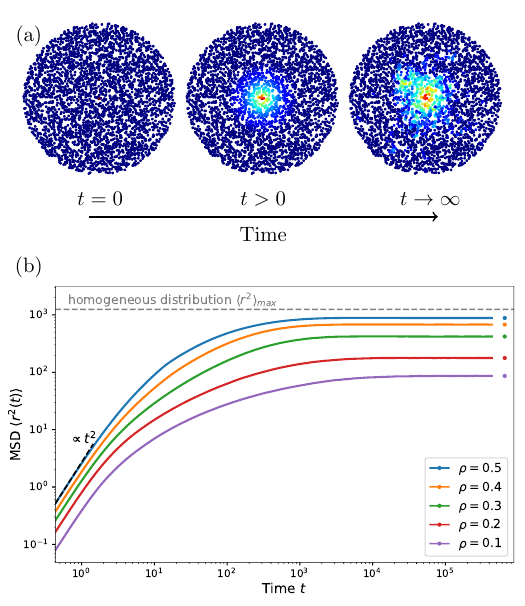}
       
    \caption{(a) Propagation of a single excitation in a Rydberg gas with $N=3000$ atoms and $\rho=0.1$. The probability $P_j(t)$ of atom $j$ for being in state $\ket{\uparrow}$ is color encoded with blue for low and red for high probability.  The spreading stops before reaching the system boundaries. Thus, the excitation is localized for $t \rightarrow \infty$.
    (b) Time dependence of the mean square displacement $\langle r^2(t) \rangle$ simulated with systems of $N=5000,4000,3000,2000,1000$ atoms for $\rho=0.5, \; 0.3, \; 0.2, \; 0.1$, respectively, with a fixed system size of $\mathcal{R} = 50$, averaged over $2000, \;2500, \; 3333, \;5000, \;10000$ random configurations. The markers on the right side represent the asymptotic late-time value $\langle r^2(t\rightarrow \infty) \rangle$ given by Eq.~\eqref{eq:p_inf}.
    }
    \label{fig: Expansion + MSD}
\end{figure}

In Fig.~\ref{fig: Expansion + MSD}\hyperref[fig: Expansion + MSD]{(b)} the transport process of a single excitation is illustrated via the MSD for multiple densities. We kept the system size constant and adjusted the number of atoms to achieve a given density. The markers on the right end of each curve represent the asymptotic late-time value of the MSD.

We observe an initial ballistic spreading [$\langle r^2(t)\rangle \propto t^2$] with a velocity that increases with density.  
This behavior is expected due to the generic quadratic onset of the evolution of populations under unitary dynamics. The velocity is determined by the typical nearest neighbor interaction strength which increases with density.
The ballistic regime is followed by a slowdown of the spreading and eventual saturation. A diffusive intermediate regime [$\langle r^2(t)\rangle \propto t$] is not recognizable. The time at which the saturation regime is reached grows with decreasing density, which is expected since the mean hopping strength scales as $\langle V_{ij}\rangle\sim\rho^{3/2}$ for $a=3$ and $d=2$.

The late-time saturation values of the MSD increase with density, consistent with less localized eigenstates for larger $\rho$. At the largest densities the MSD almost reaches the system-size limited maximal extent. In contrast, at small densities the MSD saturates far from the maximal value indicating localization, i.e., the interference induced halting of excitation transport.

\subsubsection{Size of late-time excitation distribution}
\label{sec: late time}
\begin{figure}
\includegraphics[scale=0.5]{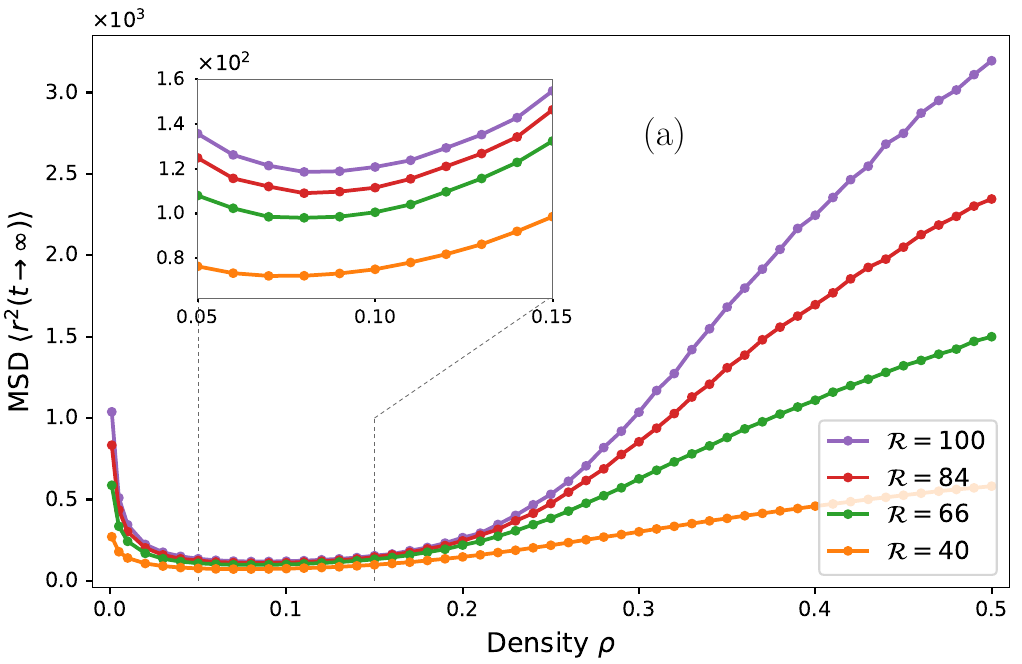}

\vspace{2mm}{\includegraphics[scale=0.5]{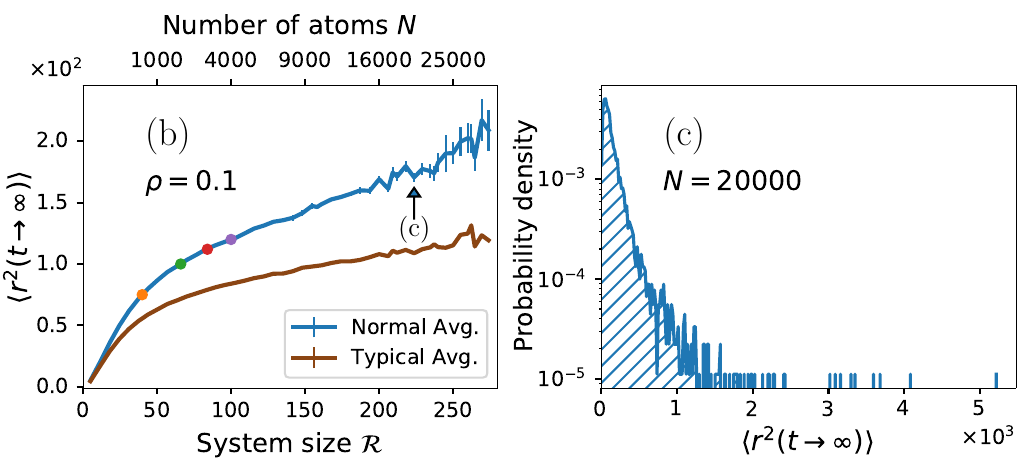}}

\vspace{2.15mm}{\caption{Density and system-size dependence of the asymptotic late-time MSD. (a) Asymptotic late-time MSD over density $\rho$ computed for four different system sizes. For each line the system radius $\mathcal{R}$ was fixed and the number of atoms was adapted to achieve the desired density $\rho$. The inset shows an enlargement of the data between $\rho=0.05$ and $\rho=0.15$. Between $3 \cdot 10^6$ disorder realizations for smallest densities and number of atoms and $500$ disorder realizations for largest densities and number of atoms were averaged. (b) Asymptotic late-time MSD over system size $\mathcal{R}$ at $\rho=0.1$. The disorder averages were performed as normal averages (blue) as well as typical, i.e., logarithmic, averages (brown). The colored markers correspond to the system sizes used in (a). The atom number was varied from $N=10$ and $N=30000$ with adapted number of disorder realizations between $2 \cdot 10^6$ and $742$. The error bars indicate the standard error of the mean over the disorder realizations. (c) Distribution of the asymptotic late-time MSD for all disorder realizations for $N=20000$.
$5177$ disorder realizations were used.
}\label{fig:MSD scaling}}
\end{figure}

\vspace{-1mm}
To decide whether a system is truly localized, i.e., to exclude that excitation transport is limited by finite system size, one needs to examine the system-size dependence of the late-time extension and show that it becomes size independent.
Figure~\ref{fig:MSD scaling}\hyperref[fig:MSD scaling]{(a)} displays the MSD for $t \rightarrow \infty$ for four different system sizes as a function of the density $\rho$. Up to a density of approximately $\rho \approx 0.2$, the MSD depends only weakly on the system size, indicating localization effects. However, we do observe a residual increase of the MSD with $\mathcal{R}$ in this regime, see inset of Fig.~\ref{fig:MSD scaling}\hyperref[fig:MSD scaling]{(a)}. 
Assuming algebraic localization of all eigenstates we can obtain a prediction for the system-size scaling of the late-time MSD. The late-time excitation distribution [Eq.~\eqref{eq:p_inf}] will inherit the power-law tails of the eigenstates. If all eigenstates had $r^{-6}$ power-law tails, one would thus expect that $n(r,\psi(t\rightarrow\infty)) = P_r(t\rightarrow\infty)\sim r^{-6}$. Assuming that the prefactor of this power-law decay is $N$-independent, this would mean that the deviation of the MSD from its infinite-$N$ value should scale as $\int_\mathcal{R}^\infty  r^2 P_r(t\rightarrow\infty) \, r\,dr  \propto \mathcal{R}^{-2}$. This expectation is, however, not met for large system sizes, where the late-time MSD does not seem to saturate but keeps increasing linearly with $\mathcal{R}$ as shown in Fig.~\ref{fig:MSD scaling}\hyperref[fig:MSD scaling]{(b)}. We attribute this effect to the presence of a small set of delocalized states present even at very low density as discussed in Sec.~\ref{sec: DOS and IPR}. The extent of these states still scales with the system size, and they do not necessarily feature power-law tails. In fact, we observed that their radial excitation density can even be nonmonotonic (see Sec.~\ref{sec: Shape}).
These delocalized states manifest as rare cases where the excitation propagates extremely far in the asymptotic late-time limit. This picture is confirmed by examining the distribution of late-time MSDs over disorder realizations in Fig.~\ref{fig:MSD scaling}\hyperref[fig:MSD scaling]{(c)}. This distribution shows a long tail towards large MSDs. The fact that this tail is extremely hard to sample is what causes the large statistical fluctuations of the late-time MSD at large system sizes visible in Fig.~\ref{fig:MSD scaling}\hyperref[fig:MSD scaling]{(b)}. We note that the linear increase of the MSD with $\mathcal{R}$ is sub-extensive as the ergodic value scales as  $\langle r^2 \rangle_{\rm max}\propto \mathcal{R}^2$.

We tested the hypothesis that the continuing increase of the late-time MSD at large system sizes is caused by a few exceptional disorder samples featuring long-range hops by additionally evaluating the typical average $\mathrm{exp} \{ \overline{ \log[\langle r^2(t\to \infty)\rangle ]}\}$ (where the overline denotes the disorder average) in Fig.~\ref{fig:MSD scaling}\hyperref[fig:MSD scaling]{(b)}, which suppresses the contribution of exceptionally large values. We found that this quantity still does not saturate at large $N$ but shows approximately power-law growth. We conclude that the fraction of delocalized states is still significant at the investigated system sizes and the observation of full localization will require even larger $N$.

At high densities the asymptotic MSD scales extensively with the system size ($\propto \mathcal{R}^2$) [see Fig.~\ref{fig:MSD scaling}\hyperref[fig:MSD scaling]{(a)}], and the extent of the excitation distribution is comparable to the system size. This reflects the fact that at densities $\rho\gtrsim 0.2$ a large fraction of eigenstates already show rather small and system-size-dependent IPRs as observed in Fig.~\ref{fig: Eigenstates}\hyperref[fig: Eigenstates]{(b)}.

An interesting feature of Fig.~\ref{fig:MSD scaling}\hyperref[fig:MSD scaling]{(a)} is the increase of the MSD for $\rho \rightarrow 0$. We attribute this to the scale invariance of the Hamiltonian for low densities. When the mean distance between nearest neighbors is much larger than the blockade
radius, atom positions are approximately uncorrelated, i.e., not affected by the Rydberg blockade radius. Thus, a configuration at one density can be described as a rescaled version of a configuration at another density, as already discussed in Sec.~\ref{sec: DOS and IPR}. 
The corresponding Hamiltonians differ from each other only by a global factor such that the eigenstates and hence also the asymptotic late-time populations remain the same. 
Thus, as rescaling lengths with $1/\sqrt{\rho}$ leaves $P_j$ invariant, the MSD $\sum r_j^2 P_j $ behaves as $\rho^{-1}$ for $\rho\rightarrow 0$.

We concluded from Sec.~\ref{sec: spectral properties} that full localization in the large $N$ limit is expected at all densities.
What Fig.~\ref{fig:MSD scaling} shows is that one cannot expect to observe this localization in the transport properties of an initially localized excitation in the sense that the average late-time distance of the excitation from its starting position as a function of system size saturates for large $N$. We still see that for low densities, at a given system size, the width of the late-time excitation distribution dynamically saturates at a value far below the one of a fully delocalized state.

The strong system-size dependence of the late-time MSD for $\rho \gtrsim 0.2$ puts stringent bounds on the density regime in which localization effects can be observed experimentally. In Rydberg atom experiments typical system sizes of $N\lesssim 3000$ have been reported in three-dimensional trap geometries \cite{Signoles2021, Orioli2018}. In quasi-two-dimensional geometries this number will typically be lower. Thus, already at moderate densities finite-size effects become inevitable due to the rapidly increasing spatial extent of the eigenstates, leading to seemingly ergodic behavior.

\vspace{-1mm}
\subsubsection{Radial shape of late-time population distribution}
\label{sec:p_late_shape}

The spatial shape of the eigenstates is expected to manifest in the asymptotic late-time distribution according to Eq.~\eqref{eq:p_inf}. We therefore study how $P_j(t\rightarrow \infty)$ decreases radially. For this we consider the radial excitation density $n(r,\psi_{t\rightarrow\infty})$, introduced in Eq.~\eqref{eq: radial density}, where instead of eigenstates we consider the asymptotic late-time state $\ket{\psi_{t\rightarrow\infty}}$ and $r$ is the distance to the cloud center, where the excitation is initially localized.

\begin{figure}
\centering
\includegraphics[scale=0.5]{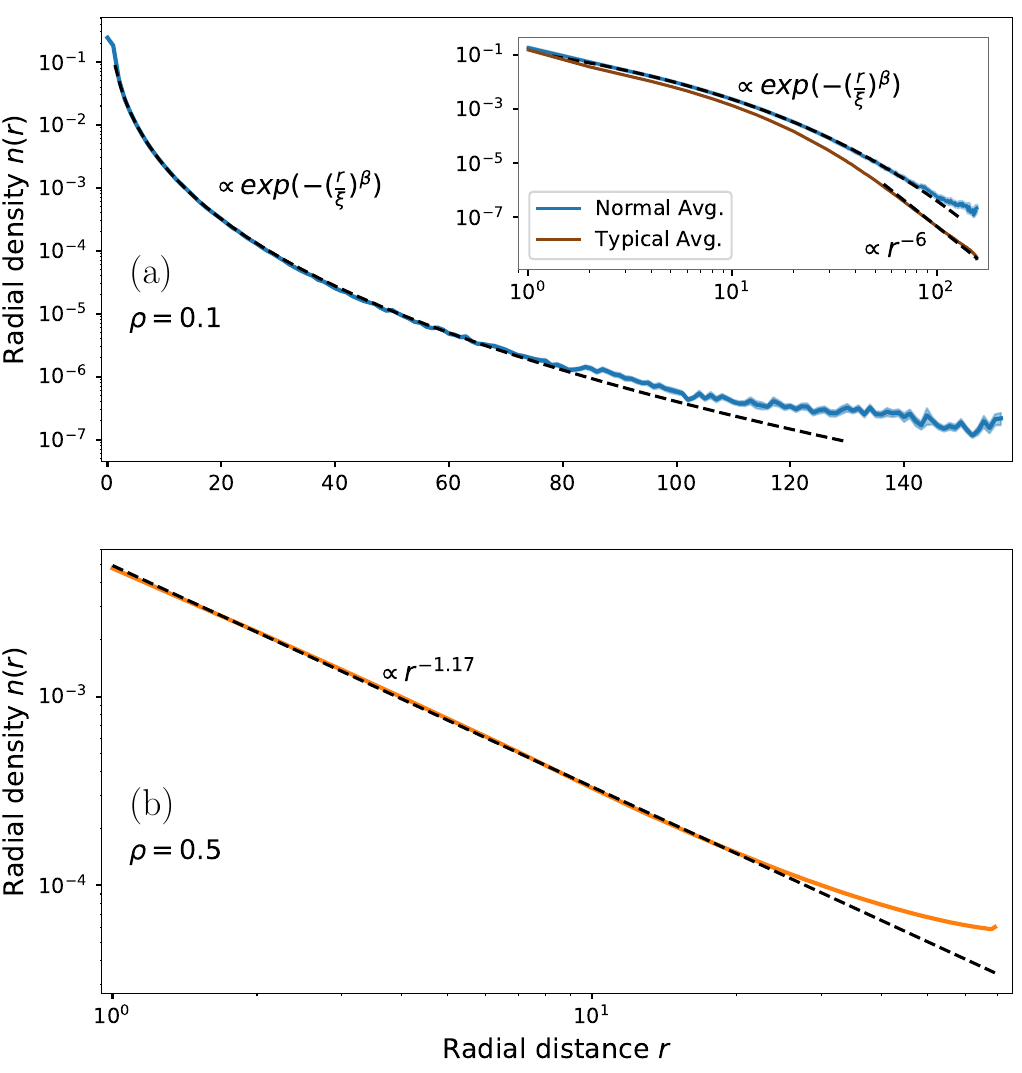}
\caption{Radial shape of the asymptotic late-time excitation density for (a) $\rho=0.1$ and (b) $\rho=0.5$ with $N=10000$ atoms and averaged over $18741$ disorder realizations for (a) and $10000$ disorder realizations for (b). The inset of (a) shows the radial density for a normal disorder average as well as for a typical average. The latter exhibits the characteristic power-law decay with $r^{-6}$.  We chose the radial bin size to be $\delta r= 1$ for both densities. The colored area shows the standard error of the mean over the disorder realizations.
Note that we used a log-linear scale in the main panel of (a), while we used a double logarithmic one in (b) and in the inset of (a).
}\label{fig: rad density}
\end{figure}

In Fig.~\ref{fig: rad density}\hyperref[fig: rad density]{(a)}, showing $n(r)$ for a density $\rho=0.1$ where all eigenstates are localized ($\mathrm{PR}\ll N$), we find that the radial density decays as a stretched exponential function, $n(r)\propto \exp[-(r/\xi)^\beta ]$ at short distances.
The stretched exponential decay is due to superposing the excitation densities of eigenstates with different decay rates of their exponential short-range behavior [according to Eq.~\eqref{eq:p_inf}].
Similarly, the power-law tails of the eigenstates (see Sec.~\ref{sec: Shape}) are expected to reflect in a power-law decay of the late-time density distribution at large $r$. However, here again the contributions of delocalized states dominate the large-$r$ tails and lead to a long tail of $n(r)$, which does not show a clear power law [see inset of Fig.~\ref{fig: rad density}\hyperref[fig: rad density]{(a)}]. Taking the typical average $\mathrm{exp}\{\overline{\log[n(r)]}\}$ allows to suppress these contributions and obtain a clear power-law tail, showing that the majority of states are algebraically localized.

For high densities, where seemingly ergodic states exist in the bulk of the spectrum, we have already seen in Fig.~\ref{fig: Expansion + MSD}\hyperref[fig: Expansion + MSD]{(a)} that the late-time MSD is close to $ \langle r^2 \rangle_{\rm max}$, which describes a homogeneous distribution. We find that for $\rho=0.5$ the late-time radial distribution fits an algebraic decay $\propto r^{-1.17}$ [dashed line in Fig.~\ref{fig: rad density}\hyperref[fig: rad density]{(b)}]. This means that the eigenstates are not perfectly ergodic, in which case a flat late-time excitation density would be expected, and is consistent with the observation made in Sec.~\ref{sec: DOS and IPR} that at high densities both localized and extended states are present.

\subsection{Effects of experimental imperfections}
\label{sec:exp_imperfections}

We have shown that at sufficiently low density $\rho$ localization effects manifest as limited spreading of an initially localized excitation, which can in principle be observed experimentally. 
However, the assumptions of unitary time evolution and homogeneous two-dimensional atom density, under which we made these observations, are never strictly fulfilled in experiments. The goal of this section is to make predictions about the possibility of observing localization effects in two-dimensional ensembles with dipolar interactions under realistic experimental conditions. For this we consider the impact of decoherence due to unavoidable experimental noise and coupling to the environment, and of nonhomogeneous and quasi-two-dimensional atomic clouds.
We restrict our discussion to the case of $\rho=0.1$, where Sec.~\ref{sec:unitary_dyn} has shown that localization effects are observable under idealized assumptions.

\subsubsection{Decoherence}

Single-particle localization is an interference effect which is destroyed by dissipative processes that reduce the coherence between different states or paths. Rydberg atoms have a finite natural lifetime of typically $100\,\mu \mathrm{s}$ (for principal quantum number $\sim 50$) after which they decay to lower lying electronic states and are susceptible to external fields leading to dephasing noise.
Observing localization is thus a matter of timescales. We therefore address the question to what extent the excitation stops spreading due to coherent localization before coherence is destroyed and incoherent diffusive dynamics takes over.

We model the effect of decoherence by means of a Lindblad master equation describing the evolution of the density matrix $\rho$, which reads
\begin{align}
\label{eq: master eq}
\dot{\rho}=-i[H,\rho] +\overbrace{\sum_k \Gamma_k \rho \Gamma_k^\dagger - \frac{1}{2} \left( \Gamma_k^\dagger \Gamma_k \rho + \rho \Gamma_k^\dagger \Gamma_k\right)}^{\text{Lindblad term }\mathcal{L}[\rho]}
\end{align}
with the jump operators $\Gamma_k$. We summarize possible decoherence effects by the jump operators $\Gamma_k=\sqrt{\gamma}\ket{k}\bra{k}$ with damping rate $\gamma$. Realistic values for $\gamma$ are $5-10\,$kHz. The Lindblad term can be simplified to
\begin{align}
\mathcal{L}[\rho]
=- \gamma \sum_{i\neq j} \rho_{ij}\ket{i}\bra{j} \,.
\end{align}
We see that including the phase-damping operators $\Gamma_k$ does not affect the diagonal elements of $\rho$, or populations, while the off-diagonal terms, or coherences, decay.
For large times any phase coherence is lost and the dynamics can be described by a classical hopping process, resulting in (sub-)diffusive behavior.
Time integration of the Lindblad master equation gives the populations $P_i(t)=\rho_{ii}(t)$ from which we calculate the MSD by $\langle r^2(t) \rangle = \sum_{j=1}^N r_j^2 P_j(t)$, as in the unitary case.

\begin{figure}
\centering
\includegraphics[scale=0.5]{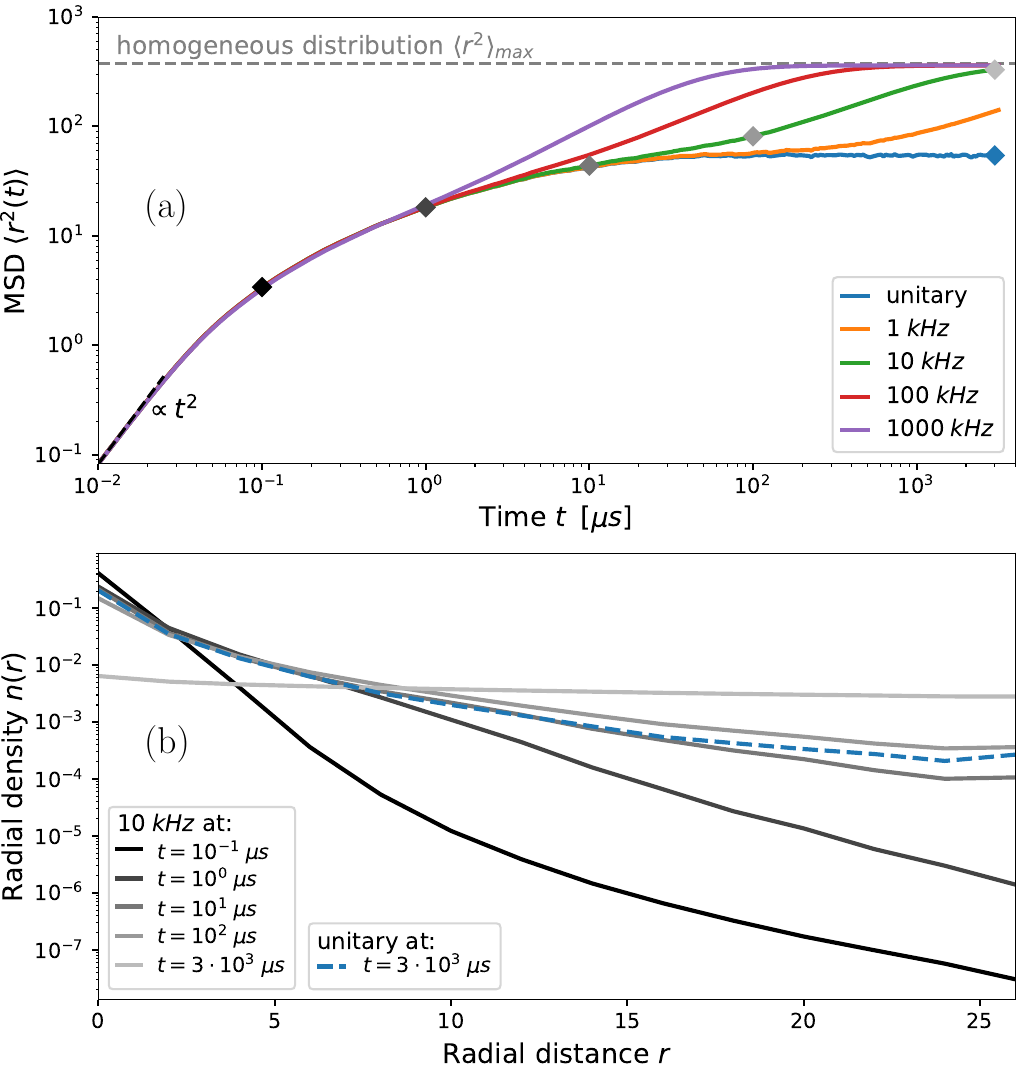}
\caption{Effects of decoherence. (a) Time dependence of the mean square displacement $\langle r^2(t) \rangle$ in a system of $N=300$ atoms with $\rho=0.1$ for different dephasing rates $\gamma =1, 10, 100, 1000\,$kHz. We averaged over $500$ disorder realizations. For better reference to typical experimental parameters we reverted the time axis to SI units here, using $C_3/2\pi=0.86\,\mathrm{GHz} \,\mu \mathrm{m}^3$ and $r_b = 5 \; \mu \mathrm{m}$. (b) Radial population density $n(r)$ for $\gamma=10\,$kHz at different evolution times showing the evolution towards a homogeneous distribution. The radial density for the unitary case (dashed) is shown for reference as well. We chose the radial bin size to be $\delta r = 2$ for all times.} \label{fig: decoherence}
\end{figure}

Figure~\ref{fig: decoherence}\hyperref[fig: decoherence]{(a)} illustrates the effect of decoherence on the MSD evolution for different damping rates $\gamma$ in a system with $\rho=0.1$ and $N=300$ atoms. For comparison, the unitary case is shown, too. 
After an initial increase, the MSD transitions into a subdiffusive regime, i.e., $\langle r^2(t) \rangle\propto t^\alpha$ with $0<\alpha<1$, where the observed $\alpha$ becomes smaller with decreasing $\gamma$. For small decoherence rates a plateau is visible before the subdiffusive increase takes over.
For very strong decoherence rates we find purely diffusive expansion with $\langle r^2(t)\rangle \propto t$ (not shown). 
At late times the fully delocalized state is approached where the MSD saturates at $ \langle r^2 \rangle_{\rm max}$.  Localization is observed if the plateau of the MSD is developed before the classical diffusive dynamics takes over.
We find that a dephasing rate of less than $\gamma = 10\,$kHz is required to clearly observe such a plateau. 
We conclude that with experimentally realistic dephasing rates of $\sim 10\,$kHz the localization plateau in the MSD will be barely visible. However, increasing the principal quantum number may allow to increase the Rydberg lifetime and also to increase the dipole-dipole interaction strength and thus accelerate unitary dynamics, allowing for a clearer observation of localization effects. 

To scrutinize the observation of full delocalization at late times, Fig.~\ref{fig: decoherence}\hyperref[fig: decoherence]{(b)} shows the radial density for $\gamma=10\,$kHz for different evolution times. At late times the distribution becomes perfectly flat as expected. This late-time behavior was observed for all dephasing rates, but takes extremely long to be established for small $\gamma$. At intermediate times, the spatial distribution comes close to the late-time distribution of the unitary case [dashed line in Fig.~\ref{fig: decoherence}\hyperref[fig: decoherence]{(b)}] showing that the late-time characteristic features of the unitary transport dynamics are still visible before diffusive transport starts to dominate.

\subsubsection{Cloud geometry}

Experimentally realistic cloud geometries will differ from the ideal case of a two-dimensional homogeneous distribution in at least two aspects. First, the cloud will typically have a Gaussian density distribution with higher density in the center than near the boundaries. Second, the system will not be strictly two-dimensional but also have a finite extent in the transverse direction, resulting in a pancake-like geometry. We have verified numerically, in the case of low density, i.e., predominantly localized eigenstates, that for a Gaussian density distribution an excitation initially localized in the center spreads very much in the same way as for a homogeneous distribution with density corresponding to the peak density of the Gaussian. Furthermore, an additional transverse density profile also does not affect the dynamics as long as the transversal width is $\sigma\lesssim r_b$.

\section{Conclusions and outlook}
\label{sec:conclusion}

We have studied spectral and eigenstate properties as well as excitation spreading in a two-dimensional power-law hopping model. The inclusion of a lower cutoff on the interatomic distances leads to a tunable strength of the off-diagonal disorder. For strong disorder almost all eigenstates are localized with power-law tails, and their properties are dominated by small localized clusters such as dimers. At finite system size decreasing the disorder strength (increasing the atomic packing fraction) leads to the appearance of extended states in the low-energy tail of the spectrum. However, a finite-size scaling analysis indicates that all states are localized asymptotically in the large $N$ limit.

Previous studies of related models have found similar indications using complementary tools and observables. We highlight that \cite{Deng2017} studied an anisotropic power-law hopping model in a two-dimensional lattice with dilute filling focusing on the level spacing ratio. Consistently with our results and with theoretical predictions \cite{deMoura2005,Kutlin2020}, they found full localization up to finite-size effects with a localization length growing quickly with filling fraction. Abumwis \textit{et al.}\ \cite{Abumwis2019b} studied a power-law hopping model in two dimensions including a blockade constraint. They focused on the eigenstate coherence $\mathcal{C}^{(n)}=\sum_{i\neq j}|c_i^{(n)}c_j^{(n)}|$ which is a measure of delocalization similar to the eigenstate PR. Interestingly, the coherence seems not to reveal the spectral features stemming from trimers and tetramers. Also, the spectral contribution of states that are more strongly localized than dimers is not obvious. In the three-dimensional anisotropic case this feature is actually absent \cite{Abumwis2019}. It would be interesting to investigate this difference between the two- and three-dimensional cases in more detail. The main finding of \cite{Abumwis2019, Abumwis2019b} is the existence of strongly delocalized states even at strong disorder in the three-dimensional case, which is heuristically understood in terms of a renormalization picture, where strongly interacting pairs are treated as being decoupled from all other atoms leading effectively to a less disordered residual system. It would be interesting to ask how this picture can be applied to understand the existence of delocalized states in three and their absence in two dimensions in the thermodynamic limit.

The observed fast growth of localization length with atomic packing density puts constraints on the regime of densities in which localization effects are observable experimentally. Limited system size and evolution time (due to decoherence effects) constrain the optimal density range above and below, respectively. At too high density the localization length easily exceeds the system size leading to seemingly ergodic behavior. At too small density dynamics are slow and excitation spreading saturates at late times where the assumption of unitary dynamics breaks down and incoherent diffusive spreading is expected. Our detailed study of these constraints allowed us to identify a packing density of $\rho = 0.1$ as a workable point. For $r_b=5\,\mu\mathrm{m}$ and $N=300$ Rydberg spins, this corresponds to a system size of $\mathcal{R}\approx137\,\mu\mathrm{m}$. According to Fig.~\ref{fig: decoherence}\hyperref[fig: decoherence]{(a)}, excitation spreading saturates at time $t\approx 20\,\mu\mathrm{s}$, requiring a decoherence rate of $\gamma \lesssim 10 \,$kHz to observe localization. All these values are within experimental reach but do not permit a large margin for varying the Rydberg spin density. Moreover, we observed that the asymptotic late-time value of the mean propagation distance of an initially localized excitation does not saturate as a function of system size, but continues to grow due to finite-size delocalized states. This precludes the observation of localization from system-size scaling of excitation transport at experimentally realistic system sizes.

Interesting future extensions of our work include the study of different power-law exponents and cloud geometries (one- and three-dimensional). Experimentally, by using two Rydberg $s$-states as spin states an XXZ model with $r^{-6}$ interactions can be realized \cite{Signoles2021}. In the single excitation sector, this corresponds to adding correlated disorder on the diagonal of the hopping Hamiltonian. In this setup isotropic interactions can also be realized in a three-dimensional geometry, in contrast to the anisotropic direct dipolar exchange interactions. Furthermore, the experimental imperfections should be modeled in more detail. While the present study models all kinds of decoherence effects as an overall dephasing process, for example, spontaneous emission and atomic-motion induced dephasing can have rather different effects and require more careful modeling. Experimentally, it may also be challenging to prepare precisely one excitation in the cloud. Thus, the case of two or more excitations should be studied, which presents a challenge to numerical methods due to the exponential growth of the Hilbert space dimension with the number of excitations.
On the theoretical side it will be interesting to apply more sophisticated analytical tools to the two-dimensional power-law hopping model. For example, one could try to extend the duality found for one-dimensional systems in \cite{Deng2018} to the two-dimensional case (see also \cite{Kutlin2020, Deng2020}). One could apply the analytical treatment based on a renormalization procedure introduced in \cite{Kutlin2020} to the present problem by including the blockade constraint. Self-consistent perturbative methods such as the locator expansion employed in \cite{Scholak2014} could be applied to predict spectral and eigenstate properties in the two-dimensional case.

\section*{Acknowledgments}

We thank A.\ Buchleitner, E.\ Carnio, T.\ Franz, I.\ Khaymovich, A.\ Signoles, S.\ Syzranov, M.\ Weidemüller, and T.\ Wellens for discussions. The authors acknowledge support by the state of Baden-Württemberg through bwHPC and the German Research Foundation (DFG) through Grant No INST 40/575-1 FUGG (JUSTUS 2 cluster), through Germany’s Excellence Strategy EXC2181/1-390900948 (the Heidelberg STRUCTURES Excellence Cluster), and through the SFB1225 ISOQUANT, Project ID 273811115.

\appendix

\section{Level spacing ratio}
\label{sec: LSR}

\begin{figure*}
\centering
\includegraphics[scale=0.5]{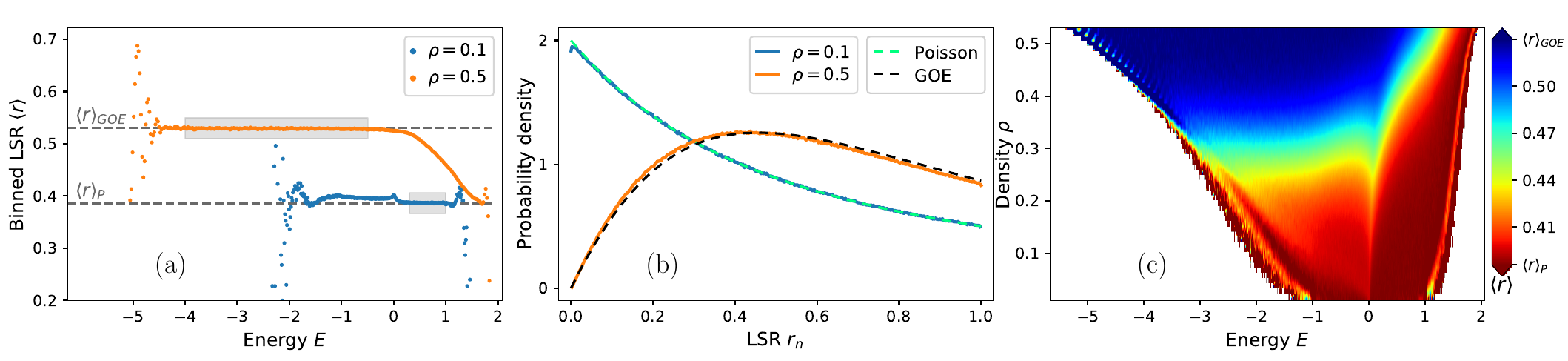}
    \caption{(a) Level spacing ratio binned into energy windows for $\rho=0.5$ and $\rho=0.1$. The energy was divided into $300$ bins which amounts to $\Delta E=0.023$. (b) Distributions of the LSR for the data from the spectral regions highlighted in gray in (a). The light green and black dashed lines show Eq.~\eqref{eq: Poi} for Poisson and Eq.~\eqref{eq: Wigner} for Wigner-Dyson. Here we divided the ratios into $500$ bins with $\Delta r=0.002$. (c) Overview of the LSR divided into several energy windows for all densities. $1000$ bins were used with $\Delta E = 0.008$. All data were obtained using for each density $N=2000$ atoms and $50000$ disorder realizations. The system parameters are the same as in Fig.~\ref{fig: Eigenstates}.}\label{fig: ratios}
\end{figure*}

In Sec.~\ref{sec: spectral properties} we characterized localization effects using eigenstate properties, in particular the IPR. An alternative indicator of localization vs.\ ergodic behavior is the statistical distribution of spacings $\delta_n = E_{n+1}-E_{n}$ between eigenenergies (sorted in ascending order), which we analyze in this appendix. The conclusions we draw from the analysis of the level statistics match the ones of Sec.~\ref{sec: spectral properties}, namely, that for finite $N$ at high densities $\rho$ (low disorder) seemingly ergodic states appear in the low-energy tail of the spectrum. However, a finite-size scaling analysis shows that states at all energies tend towards localized characteristics in the large $N$ limit.

In the following analysis we will focus on the level spacing ratio (LSR)  \cite{Deng2017, Pausch2020,  Oganesyan_2007,Atas_2013, Buijsman_2019, Kele__2019}:
\begin{align}
r_n=\frac{\min \left( \delta_n,\delta_{n-1} \right) }{\max \left( \delta_n, \delta_{n-1} \right)} \,.
\end{align}
The LSR has certain practical advantages compared to the distribution of level spacings itself. As it is a dimensionless quantity between $0$ and $1$ it allows the energy resolved characterization of the distribution of level spacings without the need of taking into account the local spectral density by unfolding the energy spectrum. An LSR of $0$ can occur only if exact level crossings are present whereas an LSR of $1$ implies that adjacent level spacings are equal.
As localized eigenstates are spatially separated and hence their eigenenergies uncorrelated, one expects a random level sequence resulting in a Poissonian level spacing distribution. In this case the distribution of $r_n$ peaks at zero, and its average is predicted as $\langle r \rangle\ind{P} \approx 0.386$. On the other hand, ergodic states extended over the whole system show spatial overlaps and consequently level repulsion. The level spacing distribution peaks at a finite value and is expected to follow a Wigner-Dyson distribution, which is the distribution found for random matrices. The model studied here has the symmetry properties of Gaussian orthogonal ensemble (GOE) for which random matrix theory predicts $\langle r \rangle\ind{GOE} \approx 0.53$.

\subsection{Dependence on energy and density}

Figure~\ref{fig: ratios}\hyperref[fig: ratios]{(a)} shows the LSR binned into energy windows for two different densities and a moderate system size with $N=2000$ atoms. At low density the level spacing ratio agrees well with $\langle r \rangle\ind{P}$ throughout the bulk of the spectrum, as expected for localized states, except for a small peak close to $E=0$. This is consistent with our analysis of the IPR [compare to Fig.~\ref{fig: Eigenstates}\hyperref[fig: Eigenstates]{(b)}]. The large variance in the eigenstate character observed around $E=0$ in the IPR does not manifest in the shown LSR plots due to the averaging over the ratios within each energy bin. 
At high density we observe that only the upper end of the spectrum shows an LSR close to $\langle r \rangle\ind{P}$. The $E<0$ tail of the spectrum is consistent with $\langle r \rangle\ind{GOE}$, corresponding to ergodic states. This again confirms the findings of Sec.~\ref{sec: spectral properties}. The fluctuations of $r_n$ at the low-energy end of the spectrum are caused by finite-size effects further discussed in Appendix~\ref{sec: Characteristics} and by poor statistics due to a small density of states.

For a more detailed test of the consistency of the observed behavior with random matrix theory predictions we study the frequency distribution of $r_n$ within selected spectral regions.
For systems with Poissonian and Wigner-Dyson level spacing distribution, respectively, the predicted distributions of $r_n$ are \cite{Atas_2013, Buijsman_2019, Kele__2019}
\begin{align}
\label{eq: Poi} 
P(r)& =\frac{2}{(1+r)^2} && \text{for Poisson,}\\
\label{eq: Wigner}
P(r)& =\frac{2}{Z}\frac{(r+r^2)^b}{(1+r+r^2)^{1+3b/2}} && \text{for Wigner-Dyson.}
\end{align}
The constants $b$ and $Z$ are ensemble dependent and take the values $b=1$ and $Z=8/27$ for the GOE.
In Fig.~\ref{fig: ratios}\hyperref[fig: ratios]{(b)} we compare the numerically obtained distributions to these predictions. We restrict to the spectral regions consistent with $\langle r \rangle\ind{P}$ and $\langle r \rangle\ind{GOE}$ for the low- and high-density case, respectively. The regions are highlighted in gray in Fig.~\ref{fig: ratios}\hyperref[fig: ratios]{(a)}.
The observed distributions coincide very well with the theoretical prediction of Eqs.~\eqref{eq: Poi} and \eqref{eq: Wigner} shown as light green and black dashed lines in the figure. The agreement is remarkable given that random matrix ensembles assume uncorrelated matrix elements while the elements of our hopping Hamiltonian \eqref{eq:HoppingHam} are correlated. Similar to the discussion in Sec.~\ref{sec: spectral properties} it is tempting to conclude from this that a transition from a fully localized phase at low density to a partly ergodic phase at high density with a mobility edge at $E=0$ exists. A careful study of system-size dependence, however, shows that the ergodic phase may disappear in the large $N$ limit.

In Fig.~\ref{fig: ratios}\hyperref[fig: ratios]{(c)} we show the binned LSR as a function of both energy and density. We observe that the LSR increases smoothly with density in all parts of the spectrum and the high-energy part of the spectrum (where the spectral bulk is located for high densities, see Fig.~\ref{fig: Eigenstates}) shows statistics of localized states even for the highest densities. The energy at which the transitions from Poisson-like to Wigner-Dyson-like statistics happens depends on density and is not necessarily at $E=0$. All these observations are consistent with what was found for the IPR in Fig.~\ref{fig: Eigenstates}\hyperref[fig: Eigenstates]{(b)}.

\subsection{System-size scaling}

\begin{figure}
    \centering
    \includegraphics[scale=0.5]{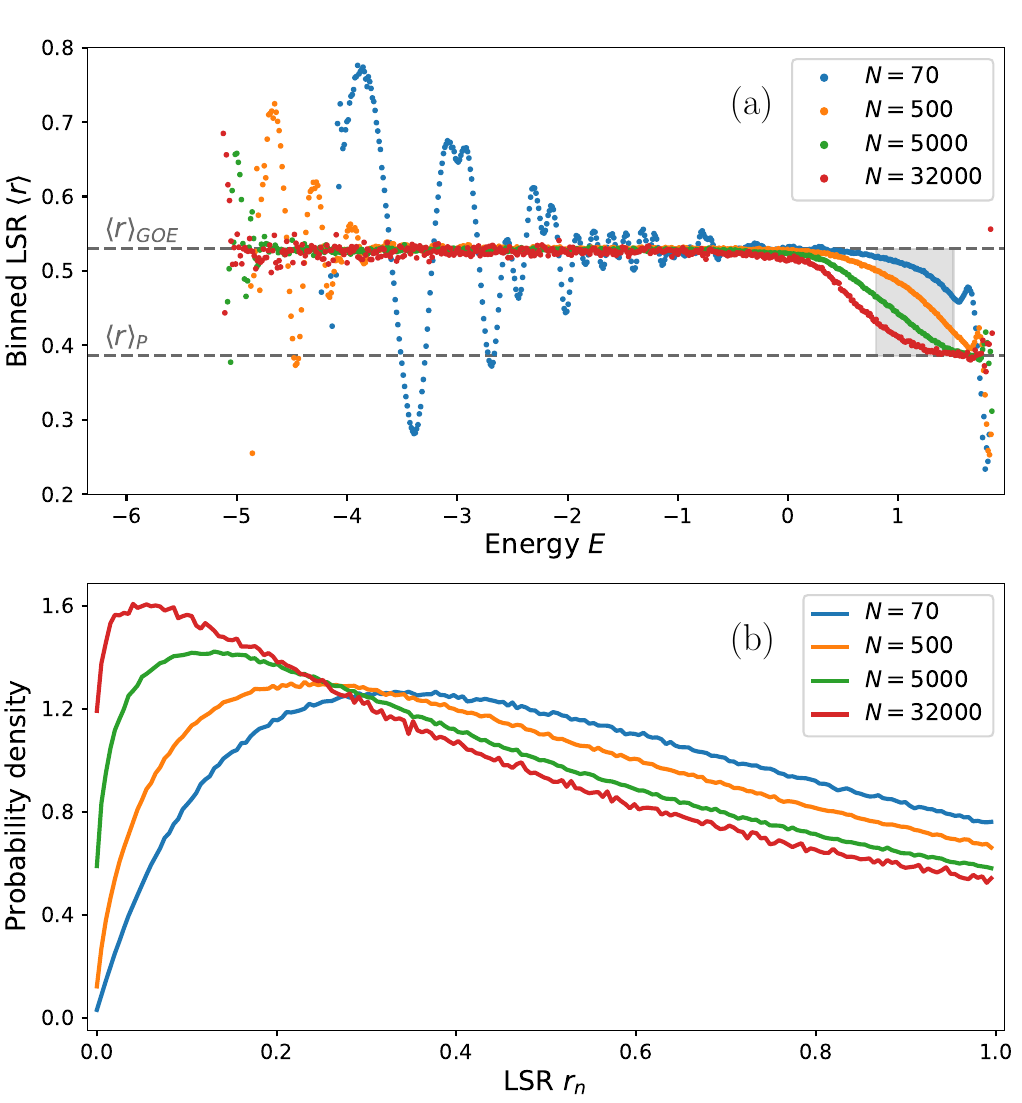}
    \caption{System-size scaling of the LSR. (a) Binned LSR for increasing system sizes with fixed density $\rho=0.5$. We used $N = 70, 500, 5000, 32000$ atoms and averaged over $5\cdot10^5, 10^5, 10^4, 200$ disorder realizations, respectively. $500$ energy bins were chosen with $\Delta E = 0.014$. (b) Distributions of the LSR of states within the energy interval marked in (a). Here we divided the ratios into $200$ bins with $\Delta r = 0.005$.}
    \label{fig: scaling ratios}
\end{figure}

To address the question whether the ergodic region persists in the large $N$ limit we now study the system-size dependence of the LSR systematically.
We calculated the LSR for multiple system sizes for a fixed density $\rho=0.5$, as shown in Fig.~\ref{fig: scaling ratios}\hyperref[fig: scaling ratios]{(a)}. For small $N$ we observe pronounced fluctuations at low $E$ which are due to finite-size effects and will be discussed in the next section. We find that the region of Poissonian level statistics at large $E$ becomes wider with increasing system size. At all energies the LSR decreases as a function of $N$, except in the regions where the LSR fluctuates due to finite-size effects (see Appendix~\ref{sec: Low energy states}). These observations suggest that all states eventually become localized and the level statistics becomes globally Poissonian at large $N$, confirming the findings of Sec.~\ref{sec: spectral properties}.

To further scrutinize this trend we examine the distribution of the LSR in a fixed energy window centered around the peak of the DOS, marked by the gray shading in Fig.~\ref{fig: scaling ratios}\hyperref[fig: scaling ratios]{(a)}, as a function of $N$. 
Figure~\ref{fig: scaling ratios}\hyperref[fig: scaling ratios]{(b)} shows that the distribution changes from Wigner-Dyson level statistics for small $N$ to Poissonian for large $N$.
Thus, while for small system size the chosen energy interval primarily contains extended states, the degree of localization grows for increasing system size. We conclude that signs of ergodicity are caused by finite-size effects and disappear in the limit of large atom number.

\section{Low-energy spectral properties}
\label{sec: Characteristics}

In this appendix we provide a microscopic understanding of the eigenstate properties at the low-energy end of the spectrum. 
For high atom densities these states can be interpreted intuitively as bound states in a quasicontinuous mean-field potential, as we show in Sec.~\ref{sec: Low energy states}.
This allows us to straightforwardly understand the origin of the oscillations in the IPR, GFD and LSR that have been observed for small system size $N$ and high density $\rho$ in Figs.~\ref{fig: Frac} and \ref{fig: scaling ratios}\hyperref[fig: scaling ratios]{(a)}, and that we study in more detail in Sec.~\ref{sec: Peculiarities}.

\subsection{Quasicontinuous mean-field picture}
\label{sec: Low energy states}

\begin{figure}[btp]
    \centering
  \includegraphics[scale=0.5]{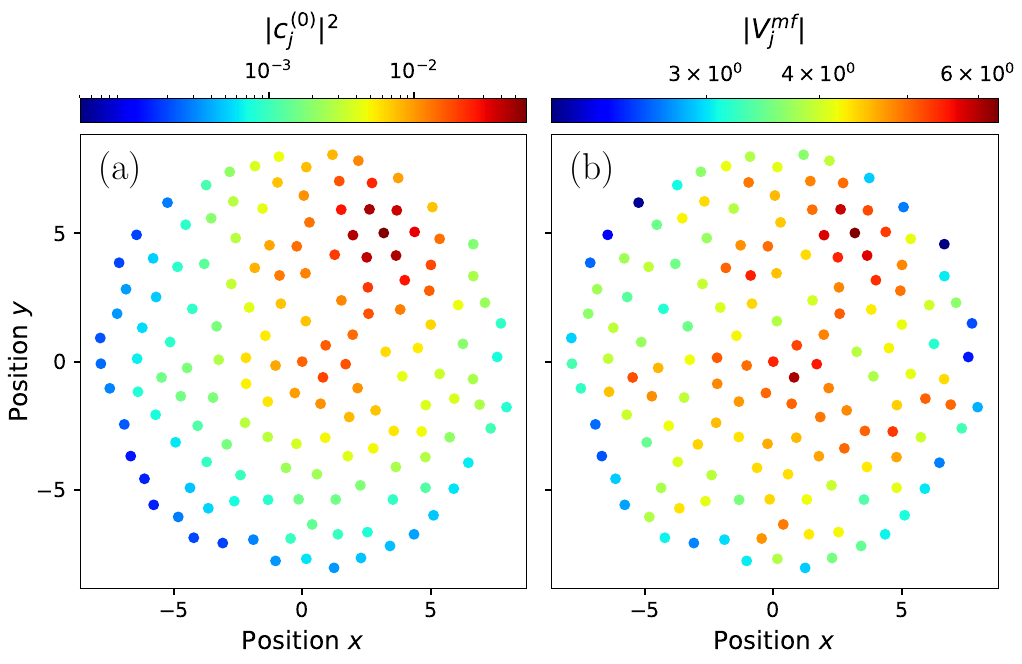}
    \caption{Comparison of the population of ground state (a) and the absolute value of the mean-field potential (b) for an example configuration with $\rho=0.5$ and $N=150$ atoms.}
    \label{fig: Mean Field}
\end{figure}

\begin{figure*}
  \vspace{-4mm}
    \centering
     \includegraphics[scale=0.5]{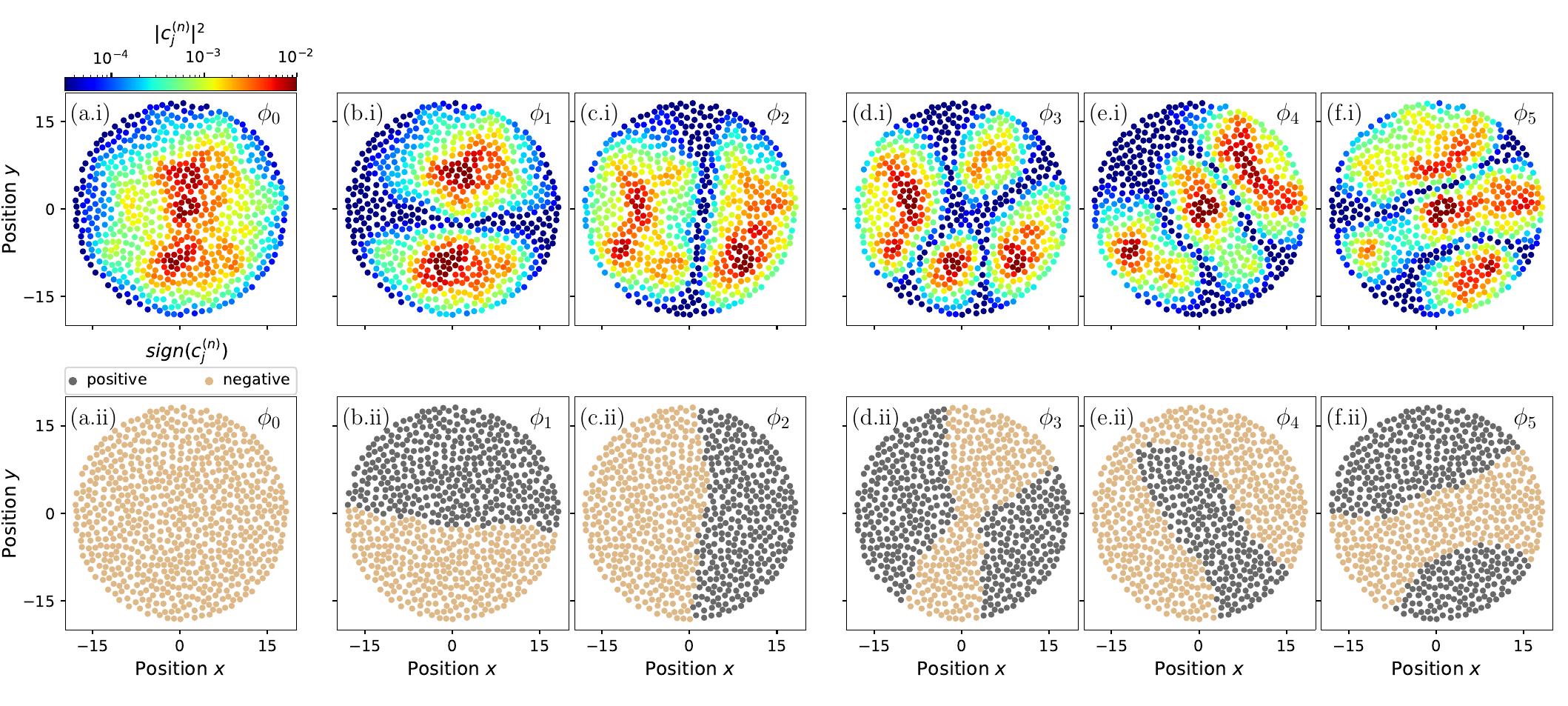}
    \caption{Populations and sign structure of the first six eigenstates for a system with $\rho=0.5$ and $N=700$ atoms. The states form groups of near-degenerate states, reminiscent of the eigenstates of the two-dimensional harmonic oscillator potential.}
    \label{fig: Harmonic Oscillator}
\end{figure*}

In this section we introduce and microscopically justify a quasicontinuous mean-field picture for explaining the properties of low-energy states. Our reasoning builds on a variational minimization of the energy expectation values [see Eq.~\eqref{eq:HoppingHam}]
\begin{equation}
\label{eq:HexpVal}
    E(\psi)=\bra{\psi}H\ket{\psi} = -\sum_{i\neq j} V_{ij} c_i^* c_j
\end{equation}
with respect to the coefficients of a general normalized state $\ket{\psi}=\sum_j c_j \ket{j}$. Since the Hamiltonian is real, we can restrict to real $c_j$. The interaction strength $V_{ij}$ is positive for all $(i,j)$ such that any pair $(c_i,c_j)$ of coefficients with opposite sign will lead to an energy increase in Eq.~\eqref{eq:HexpVal} compared to the case of equal signs. Thus, in the ground state all coefficients will have the same sign. To see how the amplitudes $|c_j|$ should be distributed in order to minimize the energy, we consider the gradients
\begin{equation}
    \left. \frac{\partial E(\psi)}{\partial c_j}\right|_{c_i=1/\sqrt{N}} = -\frac{1}{\sqrt{N}}\sum_{i\neq j}V_{ij}
\end{equation}
of the energy evaluated for homogeneously distributed coefficients $|c_j|=1/\sqrt{N}$. The negative gradient is largest for the atoms with the smallest "mean-field potential" $V^{\rm mf}_j=-\sum_{i\neq j}V_{ij}$.
This means that it is energetically favorable, compared to a homogeneous distribution of amplitudes, to enlarge the amplitudes of coefficients with low mean-field potential. Consequently, the ground state will be localized in the region of deepest mean-field potential, corresponding to regions of high local packing density of atoms. Figure~\ref{fig: Mean Field} confirms this reasoning, where an example with $N=150$ atoms and $\rho=0.5$ is shown. The excitation probability $|c_j^{(0)}|^2$ of the ground state wave function is highest in the regions of deep mean field potential, i.e., large $|V^{\rm mf}_j|$. We observe that at such high atom density $\rho$, the regions with locally highest density, i.e., deepest mean-field potential, show regular hexagonal structures of densely packed atoms at distance $r_b$ from each other.

Excited states cannot have a homogeneous sign structure as long as the ground state wave function has support on all basis states since the orthogonality of eigenstates could not be fulfilled in this case. The sign structure that introduces the smallest energy penalty is one where we have two groups of spins with opposite sign separated by a boundary line. The excitation probabilities $|c_j^{(n)}|^2$ of atoms near the boundary line are suppressed in order to minimize the energy increase due to the sign boundary. In a two-dimensional situation there are two different orthogonal states that can be created in this way. These states with different sign boundary lines are energetically close to each other. The boundary lines are expected to run through regions of lower local density as this minimize the states' energy.
With the same reasoning we can construct further states respecting orthogonality with all previously constructed states while minimizing the energy increase. From this we expect next a group of three eigenstates with three different equal-sign domains separated by boundary lines. These expectations are confirmed in Fig.~\ref{fig: Harmonic Oscillator}, which shows populations and phases for the lowest six eigenstates of a system of $N=700$ atoms and density $\rho=0.5$. The features of populations and sign structure are perfectly reminiscent of the bound states in a two-dimensional potential well. For the highest energy state (not shown), in contrast to the ground state, the signs of $c_j^{(n)}$ alternate between neighboring atoms, which can be analogously understood by variational \emph{maximization} of the energy.

To summarize, the variational construction of low-energy eigenstates leads to an intuitive quasicontinuous picture in which these states can be viewed as bound states within the potential $V^{\rm mf}$.

\subsection{Energy gaps for small system size and high density}
\label{sec: Peculiarities}

\begin{figure}[!htp]
    \centering
    \includegraphics[trim= 3pt 0 0 0]{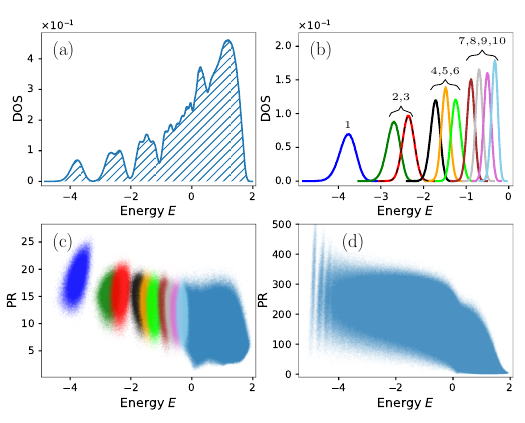}

    \caption{Oscillations in the low-energy sector for high density and small system size. (a) Energy-resolved DOS for $\rho=0.5$ and $N=30$ atoms with an accumulation over $10^6$ disorder realizations. The energy was divided into $1000$ bins with $\Delta E = 0.007$. (b) Distribution of individual eigenenergies over the random atom arrangements for each of the first ten eigenvalues showing the same data as (a). The dashed black line shows a Gaussian fit to the third eigenstate. (c) Eigenstate-resolved PRs corresponding to the data of (a). Again the colors indicate each of the ten lowest eigenstates. Here $10^5$ disorder realizations are shown. (d) PR for $\rho=0.5$ and $N=1000$ atoms with $20000$ disorder realizations.
    }\label{fig: Oscillations}
\end{figure}

We now apply the picture developed in the previous section to interpret features of the DOS and IPR in the low-energy tail of the spectrum.
Figure~\ref{fig: Oscillations}\hyperref[fig: Oscillations]{(a)} shows the DOS for high density ($\rho=0.5$) and small atom number ($N=30$). We find pronounced oscillations at low energies. By collecting the energies of the energetically lowest eigenstates resulting from different disorder realizations in separate histograms, as shown in Fig.~\ref{fig: Oscillations}\hyperref[fig: Oscillations]{(b)}, we find that the first peak represents the ground state. The gap between the ground and excited states exceeds the disorder induced fluctuations in the ground state energy. The second peak in the DOS is caused by the first \emph{and} second excited state as the gap between them is much smaller such that disorder averaging blends their spectral contributions into each other. Similarly, we can identify another gap and then a group of three states causing the third peak in the DOS. 
Using the intuition developed above we can understand these features by viewing the low-lying states as bound states within a mean-field potential (see Sec.~\ref{sec: Low energy states}). In a spherically symmetric two-dimensional potential the degeneracy of excited states increases linearly as discussed above.

We observed that the shape of the disorder-broadened spectral contributions of individual eigenstates are consistent with a Gaussian distribution. This is shown by the dashed black line in Fig.~\ref{fig: Oscillations}\hyperref[fig: Oscillations]{(b)} which corresponds to a Gaussian fit to the distribution of the third eigenstate shown in red. The width of the distributions increase with decreasing density (not shown) leading to the disappearance of the oscillations in the DOS at lower densities. 
This is expected since at lower density the interatomic distances fluctuate more strongly between different disorder realizations leading to increased fluctuations in the depth of the mean-field potential and thus larger fluctuations of the ground state energy.

Figure~\ref{fig: Oscillations}\hyperref[fig: Oscillations]{(c)} shows the participation ratios of individual eigenstates. We find that the PR is on the order of $N$ for the low-lying states, suggesting that finite-size effects are dominating their properties. Examining the PRs for larger system size, $N=1000$, shown in Fig.~\ref{fig: Oscillations}\hyperref[fig: Oscillations]{(d)}, we find that the gap between ground and excited states decreases, while states are still widely extended and the separation between different low-energy states is clearly visible. At larger system sizes, the oscillations in the DOS indeed disappear [see Fig.~\ref{fig: Eigenstates}\hyperref[fig: Eigenstates]{(a.iii)} for $N=2000$]. 

These features can be interpreted straight forwardly in our mean-field picture.
For small system sizes the ground state is extended over the whole system and is confined only by the walls of the potential  $V^{\rm mf}$ that are given by its increase at the system boundary, where a given atom simply has fewer neighbors, and thus larger $V^{\rm mf}_j=-\sum_{i\neq j} V_{ij}$. 
The observed decrease of the gap between ground and excited states with $N$ results from a wider mean-field potential, or microscopically, from a smaller energy penalty from introducing a sign boundary due to the larger spatial extent of the ground state.
In this finite-size-dominated regime the PR of the lowest lying states increases linearly with system size, which leads to an apparent fractal dimension of $\tilde{D}_2=1$, seen as a piecewise linear decrease of the IPR of the blue and orange lines at small $N$ in Fig.~\ref{fig: Frac}\hyperref[fig: Frac]{(a)}.

Let us recall here that Fig.~\ref{fig: Frac}\hyperref[fig: Frac]{(a)} showed that the ground state IPR eventually levels off and becomes independent of $N$, meaning that the ground state becomes localized. In our mean-field picture this means that the ground state is localized in a \emph{local} minimum of the mean-field potential and does not feel the potential caused by the system boundaries any more. In this regime there can also be various local minima of similar depth in the mean-field potential (corresponding to regions with densely packed atoms). Consequently one expects that energy levels become increasingly uncorrelated and tend towards Poissonian level statistics. The onset of this trend is what we indeed observe in Fig.~\ref{fig: scaling ratios}. However, showing this effect clearly, requires very large system sizes.
Our quasicontinuous mean-field picture implies that the low-energy part of the spectrum can be understood in analogy to a particle in a continuous random potential, which is known to show Anderson localization in fewer than three dimensions.

As a last interesting observation we note that the PR of the ground state, and to a lesser extent also the first few excited states, is correlated with its energy. Ground states with lower energies tend to be more strongly localized [see Figs.~\ref{fig: Oscillations}\hyperref[fig: Oscillations]{(c)} and \ref{fig: Oscillations}\hyperref[fig: Oscillations]{(d)}]. In our mean-field potential picture this is intuitively expected as stronger localization means that the state sits in a narrower and deeper potential well and thus has a lower energy.
In the microscopic view, ground states with low energies are localized on more densely packed but smaller domains than states with larger extent that experience a lower density on average and thus have higher energy.

\bibliography{refs}

\end{document}